\documentclass[aps,prd,twocolumn,nofootinbib]{revtex4-2}
\usepackage{graphicx}
\usepackage[utf8]{inputenc}
\usepackage{placeins}
\usepackage{amssymb, amsmath}
\usepackage{tabularx}
\usepackage[normalem]{ulem}
\usepackage{pbox}
\usepackage{float}
\usepackage[inline]{enumitem}
\usepackage{multirow}
\usepackage{array}
\newcommand{\Ms}{M_{\odot}}

\usepackage[dvipsnames, usenames]{xcolor}

\begin{document}

\title{Stiffness corrections for \textit{f}-mode neutron star universal relations }
\author{Isaac G. Chu and Carolyn A. Raithel}
\affiliation{Department of Physics and Astronomy, Swarthmore College, Swarthmore PA 19081, USA}

\begin{abstract}
Excitations of the neutron star (NS) fundamental mode ($f$-mode) during a binary NS inspiral or from a glitching pulsar can produce gravitational waves (GWs) that will be detectable with the next generation of detectors, opening the door to GW asteroseismology of neutron stars. There exist approximately equation-of-state (EOS)-insensitive, or ``universal," relations that relate the $f$-mode to other NS properties. These universal relations have typically been identified with relatively small, selective EOS samples. To more thoroughly explore the parameter space, we examine the \textit{f}-modes for a large sample of piecewise-parametric EOSs by solving the linearized fluid perturbation equations for the \textit{f}-mode frequency and damping time in full general relativity. We utilize our sample to evaluate the EOS-sensitivity of previously-proposed universal relations (URs) for both the \textit{f}-mode frequency and damping time. We find that URs that depend on combinations of only NS mass and radius have significant residuals, as previously reported; but that these residuals are strongly correlated with the overall stiffness of the EOS. We account for this residual dependence on the EOS by introducing linear correction terms that depend on the characteristic radius predicted by the EOS and the slope of the mass-radius curve at intermediate masses. To assess the impact of our new correction terms on observables, we consider the case of GW emission due to a glitch in a Vela-like pulsar and demonstrate that the predicted signal-to-noise ratio for such an event is over/under-estimated by up to $\sim10$\% when using an existing, single-parameter UR, but is recovered nearly exactly with our improved 3-parameter UR. In contrast, we find that URs between the $f$-mode and tidal deformability exhibit only minimal EOS-dependent correlations in their residuals and overall remain highly EOS-insensitive.
\end{abstract}

\maketitle

\section{Introduction}
Gravitational waves (GWs) from binary neutron star (NS) mergers offer a powerful way of probing the currently-unknown equation of state (EOS) of ultra-dense matter. The first such event, GW170817 \cite{TheLIGOScientific:2017qsa},  provided novel constraints on the adiabatic tidal deformability, $\Lambda$, of the merging NSs \cite[e.g.,][]{Baiotti:2019sew,Raithel:2019uzi,GuerraChaves:2019foa,Chatziioannou:2020pqz}, which affects the phase of the waveform at intermediate frequencies ($\lesssim 800$~Hz). Later in the inspiral, dynamical tides (i.e., with a frequency-dependent tidal response) can arise, as the NS's internal oscillation modes approach resonance with the orbital frequency. Among these modes, the fundamental quadrupolar mode, or $f$-mode, dominates the dynamical part of the tide \cite{Andersson:2019dwg} and can lead to significant phase effects in the GW waveform \cite{Hinderer:2016eia,schmidthinderer,Ma:2020rak}. The first constraints on the NS $f$-mode frequency from GW170817 are not informative, but future detector networks will be able to measure $f$-mode frequencies to within 10s of Hz \cite{Pratten:2019sed}. To account for these effects, a growing number of waveform models are being developed to include dynamical tides \cite{Hinderer:2016eia,Steinhoff:2016rfi,schmidthinderer,Steinhoff:2021dsn,Gamba:2022mgx,Abac:2023ujg}. Already by the fifth LIGO observing run, neglecting $f$-mode contributions during GW parameter estimation can lead to large systematic biases in the inferred tidal deformability, corresponding to errors in the inferred neutron star radius of $\mathcal{O}(1)$km  \cite{Pratten:2021pro}. 

Following a binary NS merger, post-merger GWs from the hot, rapidly-rotating, massive neutron star remnant may provide additional insight. In particular, the dominant fluid oscillation mode powering these post-merger GWs has been associated with the quadrupolar $f$-mode of the remnant \cite{Stergioulas:2011gd,Bauswein:2015vxa,Bauswein:2015yca,Rezzolla:2016nxn}; and the peak frequency of the post-merger GWs is strongly correlated with the $f$-mode frequency of a static, isolated NS \cite{Chakravarti:2019sdc,Lioutas:2021jbl}. 
Discrepancies between EOS constraints obtained from the inspiral and from the post-merger \textit{f}-mode frequencies can indicate the emergence of new degrees of freedom or phase transitions in the remnant  \cite[e.g.,][]{Bauswein:2018bma,Most:2018eaw,Weih:2019xvw,Bauswein:2020ggy,Blacker:2020nlq,Raithel:2022orm,Breschi:2023mdj}. 

Considering systems beyond binary NS mergers, GWs can also arise from excitations of a NS's \textit{f}-mode due to a pulsar glitch. Current GW observatories may already have the sensitivity to observe GWs from Vela-sized glitches in nearby pulsars, while the next-generation of GW detectors might observe a significant number of such events \cite{2019GCN.26243....1K,ho_primary}.

Measurements of the $f$-mode from such sources offer a new possible probe of NS structure, either through direct inference of EOS parameters \cite[e.g.,][]{Pradhan:2023zor,Pradhan:2023zmg} or through so-called ``universal" relations (URs). Similar to the well-known I-Love-Q relations  \cite{iloveq}, the $f$-mode URs are a family of approximately EOS-insensitive, empirical relations that relate the $f-$mode frequency to other NS parameters, such as the stellar compactness, moment of inertia, or tidal deformability  \cite{andersson,Tsui:2004qd,Lau:2009bu,Chan:2014kua,Chirenti:2015dda, Sotani:2021kiw, Sotani:2021nlx, Kunjipurayil:2022zah, Pradhan:2022rxs, Zhao:2022tcw}.

The existence of these URs reveals the quasi-incompressibility of neutron stars in certain slow physical processes, such as during adiabatic tidal deformations or $f$-mode oscillations \cite{Chan:2014kua,Katagiri:2025qze}. The $f$-mode URs are also of important practical use, providing a method for bypassing the higher computational cost to calculate the \textit{f}-mode in favor of more straightforward TOV calculations. Indeed, many studies that estimate the observability of dynamical tide effects or $f$-mode glitches use the $f$-mode URs in this way \cite[e.g.,][]{Pratten:2021pro,williams,Wilson:2024tcc}. In addition, various versions of NS URs can be used to measure the Hubble constant from GW data  \cite{PhysRevD.104.083528}; break degeneracies in GW parameter estimation \cite{Chatziioannou:2018vzf,Xie:2022brn}; and test theories of modified gravity \cite[e.g.,][]{iloveq}. For all of these purposes, well-calibrated URs are essential. 

To quantify the degree of universality in the $f$-mode relations, early works identified UR fits and quantified their residuals using broad but relativity small samples of $\mathcal{O}(10-20)$ tabulated or polytropic EOSs \cite{andersson,Tsui:2004qd,Lau:2009bu,Chan:2014kua,Chirenti:2015dda}. Although sufficient to characterize the initial URs, this sample size is likely too small to account for the full range of physical possibilities allowed by current uncertainties in the EOS. To more fully explore the pressure-density parameter space, more recent studies have introduced samples of $\mathcal{O}$(1000s) of parametric EOSs, to recalibrate existing $f$-mode URs and, in some cases, reveal new ones \cite{Pradhan:2022rxs,Zhao:2022tcw,PhysRevD.103.063036}.

In this paper, we examine several previously-posed URs for both the \textit{f}-mode frequency and its damping time. To perform a thorough statistical analysis, we use a sample of 30,720 randomly-generated, piecewise-polytropic EOSs to investigate a large portion of the EOS parameter space. In particular, we focus on the average density, effective compactness, and tidal deformability relations for the frequency, and look at damping time relations dependent on compactness and tidal deformability.

We show that several of these existing, single-parameter relations exhibit residual EOS sensitivity (i.e., scatter in the UR) which is significantly correlated with the characteristic NS radius ($R_{1.4}$) and the approximate slope of the mass-radius relation ($R_{1.4}/R_{1.8}$; where $R_{\alpha}$ is the predicted radius of a NS with mass $\alpha$).  These two parameters are associated with the EOS pressure at $\sim$2 and 3-4$\times$ the nuclear saturation density  \cite{Lattimer:2000nx,ozelpsaltis}, and hence can be viewed as macroscopic proxies for the underlying EOS stiffness. In choosing these parameters, we are also motivated by Ref.~\cite{Raithel:2022orm}, who demonstrated that the UR between the peak frequency of the post-merger GWs and the NS radius at fixed mass breaks down as the mass-radius slope is systematically varied. Similarly, Ref.~\cite{Tan:2021nat} showed that the binary-Love UR \cite{iloveq}, which was previously thought to be fully universal, also has an implicit dependence on the mass-radius slope. 

Overall, we find the largest 1$\sigma$ residuals (up to $\sim4\%$) in the URs that relate either the real or imaginary components of the $f$-mode frequency to combinations of the NS mass and radius, and we find that the residual correlations with $R_{1.4}$ and the mass-radius slope are the strongest for these relations ($R^2\gtrsim0.7$). Consistent with previous work, we find that relations between the $f$-mode and the tidal deformability or moment of inertia are tighter (sub-percent residuals); but, even with the smaller scatter of the moment of inertia relation, there are significant residual correlations with the EOS stiffness. In contrast, relations with the tidal deformability have both small residuals and minimal dependence on the EOS stiffness within those residuals.

These residual correlations with EOS stiffness imply that some $f$-mode URs will fail to accurately represent the stiffest or softest models and will potentially introduce a \textit{systematic} bias into their predictions. To address this, we introduce correction terms to the URs that depend on $R_{1.4}$ and $R_{1.4}/R_{1.8}$, which reduce the average 1-$\sigma$ residuals for the full sample (i.e., all EOSs across all masses) and for a sub-sample of 1.4$\Ms$ NSs  by as much as $\sim 60\%$ and $\sim85\%$, respectively. The addition of these stiffness correction terms thus provides a straightforward way to improve the universality of $f$-mode URs that depend on combinations of the NS mass and radius or on the moment of inertia.

The rest of this paper is laid out as follows. In Section \ref{sec:methods}, we detail the numerical methods used to calculate the \textit{f}-mode frequencies. In Section \ref{sec:results}, we outline the EOS sample used and describe our results for all of the URs we consider. In Section \ref{sec:glitches}, we investigate how using the original, single-parameter $\sqrt{M/R^3}$ UR can influence estimates of the SNR of gravitational waves from pulsar glitches, as opposed to using the updated, multi-parameter UR proposed here. In Section \ref{sec:discussion}, we discuss our results and their implications. Appendix \ref{sec:appendixdamp} contains additional damping time relations, while Appendix \ref{sec:appendixparams} contains tables of best-fit parameters for all URs considered.

\section{Quasinormal Oscillations in General Relativity} \label{sec:methods}
We consider a quasinormal mode decomposition of NS oscillations in full general relativity \cite{Thorne,Detweiler:1985zz}. Following the treatment in Refs.~\cite{Thorne,Kunjipurayil:2022zah}, we examine an even-parity perturbation of the Regge-Wheeler metric. The corresponding line element (using geometrized units $c=G=1$) for the $l,m$th mode with complex frequency $\omega$ is as follows:
\begin{multline}
    ds^2 =  -e^{\nu(r)}[1+r^lH_0(r)e^{i\omega t}Y_{lm}(\phi,\theta)]dt^2 \\+ e^{\lambda(r)}[1-r^lH_0(r)e^{i\omega t}Y_{lm}(\phi,\theta)]dr^2\\+[1-r^lK(r)e^{i\omega t}Y_{lm}(\phi,\theta)]r^2d\Omega^2 \\-2i\omega r^{l+1}H_1(r)e^{i\omega t}Y_{lm}(\phi,\theta)dtdr,
\end{multline}
where the coefficient functions are defined in the subsection below. In what follows, we solve for the $l=2$ ($f$-)mode.

\subsection{The Stellar Interior}
The metric functions $\lambda(r)$ and $\nu(r)$ are determined by the Tolman-Oppenheimer-Volkoff equations, which govern relativistic stellar structure, according to
\begin{equation}
    e^{\lambda(r)} = \frac{1}{1-2b(r)},
\end{equation}
\begin{equation}
    \frac{d\nu}{dr} = -\Big(\frac{2}{P+\varepsilon}\Big)\frac{dP}{dr}, 
\end{equation}
where $b(r) = m(r)/r$, $m(r)$ is the mass enclosed by radius $r$, $P$ is the pressure, and $\varepsilon$ is the energy density. $\nu$ obeys the boundary condition $e^{\nu(R)} = 1-2b(R)$ at the edge of the star, $R$. 

The functions $H_0(r)$, $H_1(r)$, and $K(r)$ describe the radial variation of the metric perturbation, while the spherical harmonics $Y_{lm}$ describe the angular dependence. The complex eigenfrequency $\omega$ has a real part corresponding to the oscillation frequency and an imaginary part corresponding to 1/$\tau$, where $\tau$ is the damping time of the oscillation.

In addition to the metric perturbation, there is also a corresponding perturbation to the fluid itself, which can be written as a Lagrangian displacement vector
\begin{align}
    &\xi^r = r^{l-1}e^{-\lambda/2}We^{i\omega t}Y_{lm} \\
    &\xi^\theta = -r^{l-2}Ve^{i\omega t}\partial_\theta Y_{lm} \\
    &\xi^\phi = -\frac{r^{l-2}}{\sin^2\theta}Ve^{i\omega t}\partial _\phi Y_{lm},
 \end{align}
where $W$ and $V$ are the fluid perturbation amplitudes.
$X$ is a function, related to a Lagrangian pressure perturbation, defined as follows
\begin{equation}
\label{eq:deltaP}
    \Delta P = -r^le^{-\nu/2}Xe^{i\omega t}Y_{lm}.
\end{equation}
By using Einstein's equation to relate $X$ to $H_0$, $H_1$, and $K$, one of the original five degrees of freedom (three from the metric, 2 from the fluid) can be eliminated. Refs.~\cite{lindblomDOF,lindblomseminal} elect to eliminate $H_0$ to avoid a singularity in the system of ODEs for a certain range of frequencies, leaving a system of four differential equations for $H_1$, $K$, $W$, and $X$:  
\begin{equation}
\begin{split}
    r\frac{dH_1}{dr} = &-[l+1+2be^\lambda+4\pi r^2e^\lambda(P-\varepsilon)]H_1 \\ &+e^\lambda[H_0+K-16\pi(\varepsilon+P)V],
\end{split}
\end{equation}
\begin{equation}
    \begin{split}
        r\frac{dK}{dr} = &H_0 + (n+1)H_1 \\ &+[e^\lambda Q-l-1]K-8\pi(\varepsilon+P)e^{\lambda/2}W,
    \end{split}
\end{equation}
\begin{equation}
    \begin{split}
        r\frac{dW}{dr} = &-(l+1)[W+le^{\lambda/2}V] \\ &+r^2e^{\lambda/2}\Bigg[ \frac{e^{-\nu/2}X}{(\varepsilon+P)c^2_{ad}}+\frac{1}{2}H_0+K\Bigg],
    \end{split}
\end{equation}
\begin{equation}
    \begin{split}
        r\frac{dX}{dr} = &-lX+\frac{1}{2}(\varepsilon+P)e^{\nu/2} \Bigg\{ (1-e^\lambda Q)H_0 \\ &+(r^2\omega^2e^{-\nu}+n+1)H_1+(3e^\lambda Q-1)K\\ &-4(n+1)r^{-2}e^\lambda QV-2\Big[ \omega^2e^{\lambda/2-\nu} \\ &+4\pi (\varepsilon+P)e^{\lambda/2} -r^2\frac{d}{dr}\Big(\frac{e^{\lambda/2}Q}{r^3} \Big)\Big]W \Bigg\},
    \end{split}
\end{equation}
where $Q = b+4\pi r^2P$, $n = (l-1)(l+2)/2$, and $c^2_{ad}$ is the adiabatic sound speed within the star. We approximate $c^2_{ad}$ with the equilibrium sound speed $c^2_{eq} = dP/d\varepsilon$ throughout this paper \cite{Kunjipurayil:2022zah}.

The quantities $H_0$ and $V$ can then be solved for algebraically as a linear combination of $H_1$, $K$, $W$, and $X$:
\begin{multline}
    H_0 = \Big\{8\pi r^2e^{-\nu/2}X - \Big[ (n+1)Q-\omega^2r^2e^{-(\nu+\lambda)}\Big]H_1\\+[n-\omega^2r^2e^{-\nu}-Q(e^\lambda Q-1)]K\Big\}(2b+n+Q)^{-1},
\end{multline}
\begin{equation}
    V = \Bigg[ \frac{X}{\varepsilon+P}-Qr^{-2}e^{(\nu+\lambda)/2}W-\frac{1}{2}e^{\nu/2}H_0 \Bigg]\omega^{-2}e^{\nu/2}.
\end{equation}

To solve this system of differential equations, we apply the following boundary conditions:
\begin{equation}
    W(0) = 1,
\end{equation}
\begin{equation}
    \begin{split}
        X(0) = &(\varepsilon_0+P_0)e^{\nu_0/2} \\ &\Bigg\{ \Bigg[ \frac{4\pi}{3}(\varepsilon_0+3P_0)-\frac{\omega^2}{l}e^{-\nu_0} \Bigg] W(0) + \frac{1}{2} K(0)\Bigg\},
    \end{split}
\end{equation}
\begin{equation}
    H_1(0) = \frac{lK(0) + 8\pi (\varepsilon_0+P_0)W(0)}{n+1},
\end{equation}
\begin{equation}
    X(R) = 0.
\end{equation}
To satisfy the final boundary condition, we first perform two trial solutions of the system from $r=0$ to the edge of the star with $K(0) = \pm (\varepsilon_0 +P_0)$, and then perform a linear interpolation to find the initial value of $K$ that will yield $X(R) = 0$ (implying no variation in pressure at the surface of the star; see eq. \ref{eq:deltaP}). We also note that $H_0(0) = K(0)$ by construction. 

To solve for $\omega$ at a given central density, we start with a trial value (e.g. 1.6 kHz), then perform the above process using a 5th-order Runge-Kutta method, with relative error tolerance held at $10^{-6}$.

Alongside these \textit{f}-mode, mass, and radius calculations, we also calculate the moment of inertia \cite{PhysRevD.61.024009} and the adiabatic tidal deformability of the NS \cite{Hinderer_2008}.

\subsection{The Exterior Solution}
Outside of the stellar surface, the only remaining perturbations are those of the metric. Their behavior is governed by Zerilli's equation
\begin{equation} \label{zerillidiff}
    \frac{d^2Z}{dr^{*2}} = (V_Z-\omega^2)Z,
\end{equation}
where $r^* = r+2M\ln (\frac{r}{2M}-1)$. The effective potential $V_Z$ is defined as \cite{zerilli_fackerell}:
\begin{equation}
    V_Z = (1-2b)\frac{2n^2(n+1)+6n^2b+18nb^2+18b^3}{r^2(n+3b)^2}.
\end{equation}
Zerilli's function $Z$ is defined by the following linear transformation:
\begin{equation} \label{matchingzerilli}
    \begin{split}
        \begin{pmatrix}
            K(r) \\ H_1(r)
        \end{pmatrix} &= \begin{pmatrix}
            g(r) & 1 \\ h(r) & k(r)
        \end{pmatrix} \begin{pmatrix}
            Z(r^*)/r \\ dZ(r^*)/dr^*
        \end{pmatrix}, \\ g(r) &= \frac{n(n+1)+3nb+6b^2}{n+3b}, \\ 
        h(r) &= \frac{n-3nb-3b^2}{(1-2b)(n+3b)}, \\
        k(r) &\equiv \frac{dr^*}{dr} = \frac{1}{1-2b}.
    \end{split}
\end{equation}
After performing the final solve of the stellar interior, we use this transformation to find the initial values of $Z$ and $\frac{dZ}{dr^*}$. We then integrate from the stellar surface to $r>100$ km, so that we obtain the asymptotic limit of $Z$. In this far-field limit, $Z$ is decomposed into incoming and outgoing gravitational waves, according to:
\begin{equation}\label{asymptoticzerilli}
    \begin{split}
        \begin{pmatrix}
            Z(\omega) \\ dZ/dr^*
        \end{pmatrix} &= \begin{pmatrix}
            Z_-(\omega) & Z_+(\omega) \\ dZ_-/dr^* & dZ_+/dr^*
        \end{pmatrix} \begin{pmatrix}
            A_-(\omega) \\ A_+(\omega)
        \end{pmatrix}, 
    \end{split}
\end{equation}
where
\begin{equation}
   \begin{split}      
        \\ Z_- &= e^{-i\omega r^*}\Big[ \alpha_0+\frac{\alpha_1}{r}+\frac{\alpha_2}{r^2} + \mathcal{O}(r^{-3}) \Big], \\
        \frac{dZ_-}{dr^*} &= -i\omega e^{-i\omega r^*}\Big[ \alpha_0 + \frac{\alpha_1}{r}  \\ &+\frac{\alpha_2 - i\alpha_1(1-2b)/\omega}{r^2} +\mathcal{O}(r^{-3}) \Big], \\
        \alpha_1 &= \frac{-i(n+1)\alpha_0}{\omega}, \\
        \alpha_2 &= \frac{[-n(n+1) + iM\omega(3+6/n)]\alpha_0}{2\omega^2}.
    \end{split}
\end{equation}
Here, $A_+(\omega)$ and $A_-(\omega)$ are the amplitudes of ingoing and outgoing radiation, respectively, and $Z_+$ is the complex conjugate of $Z_-$.

\begin{figure}[ht]
\centering
\includegraphics[width=0.95\linewidth]{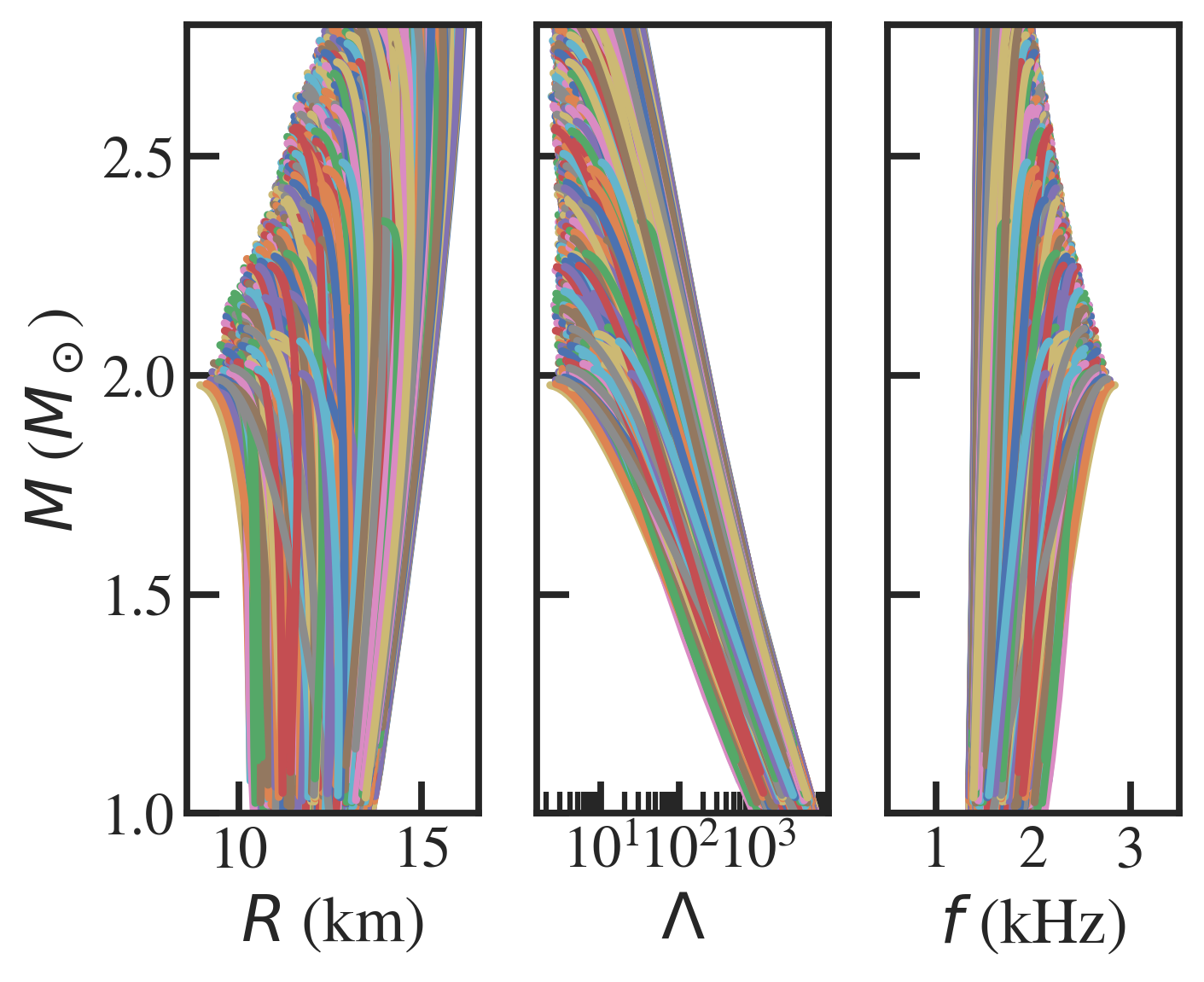}
\caption{\label{fig:eossample} Mass vs. radius, mass vs. adiabatic tidal deformability $\Lambda$, and mass vs. \textit{f}-mode frequency curves for the 30,720 EOSs in the parametric EOS sample.}
\end{figure}

\begin{figure}[ht]
    \centering
    \includegraphics[width=0.85\linewidth]{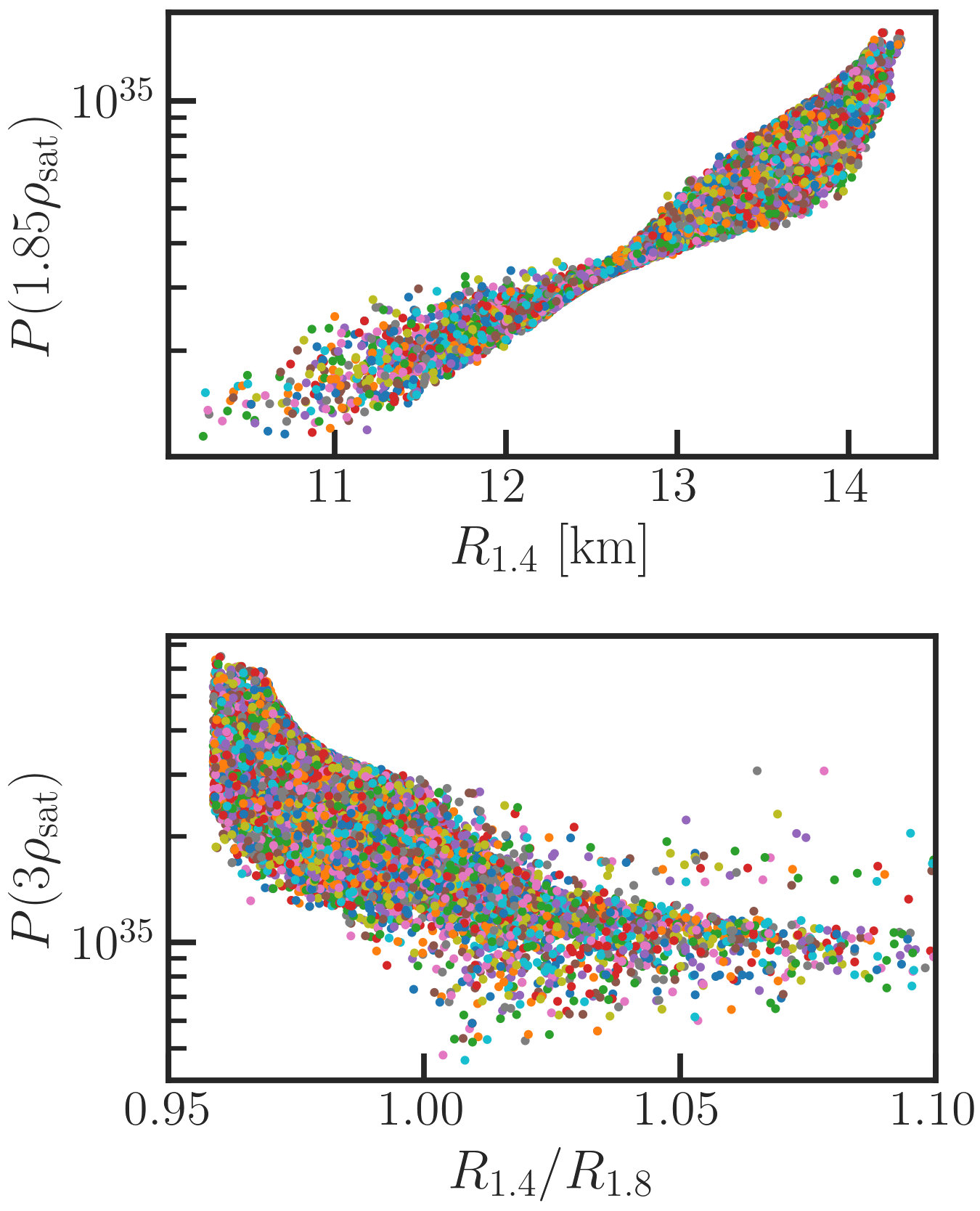}
    \caption{Top: Correlation between the pressure at 1.85$\rho_{\rm sat}$ and the radius of a 1.4$\Ms$ NS, $R_{1.4}$, for our piecewise-polytropic sample of EOSs. Bottom: Correlation between the pressure at higher densities (here, 3$\rho_{\rm sat}$) and the ratio $R_{1.4}/R_{1.8}$, which is proportional to the mass-radius slope at intermediate masses. }
    \label{fig:R14_slope}
\end{figure}
\subsection{Solving for \textit{f}-modes}
Finally, we solve the eigenvalue problem for the complex oscillation frequency $\omega$. This amounts to imposing the condition that gravitational radiation must be strictly outgoing in the far-field limit, i.e. $A_+(\omega) = 0$. To perform this calculation, we search between $\omega_{\rm min} = 644$ Hz and $\omega_{\rm max}=3221$ Hz for the value of $\omega_{\rm crit}$ where Re$(A_+(\omega))$ = 0, to an accuracy of 1 Hz. This range conservatively brackets the expected values of $\omega$ expected for astrophysical NSs. Then, we take 30 sets of ($\omega$,$|A_+(\omega)|^2$) values surrounding $\omega_{\rm crit}$ and we fit a quadratic \cite{lindblomseminal} to these values. By finding the root of this quadratic fit function, we thus solve for the real and imaginary components of $\omega$ \cite{lindblomseminal}.

This algorithm solves for the $f$-mode eigenfrequency; additional eigenfrequencies (e.g. p-modes and g-modes) can be solved for in a similar way, except the radial perturbation functions have a non-zero number of turnovers (nodes) \cite{Kunjipurayil:2022zah}. In the case of the \textit{f}-mode, the perturbation functions have 0 nodes.

\section{Results} \label{sec:results}
\begin{figure*}[ht]
\centering
\includegraphics[width=0.85\textwidth]{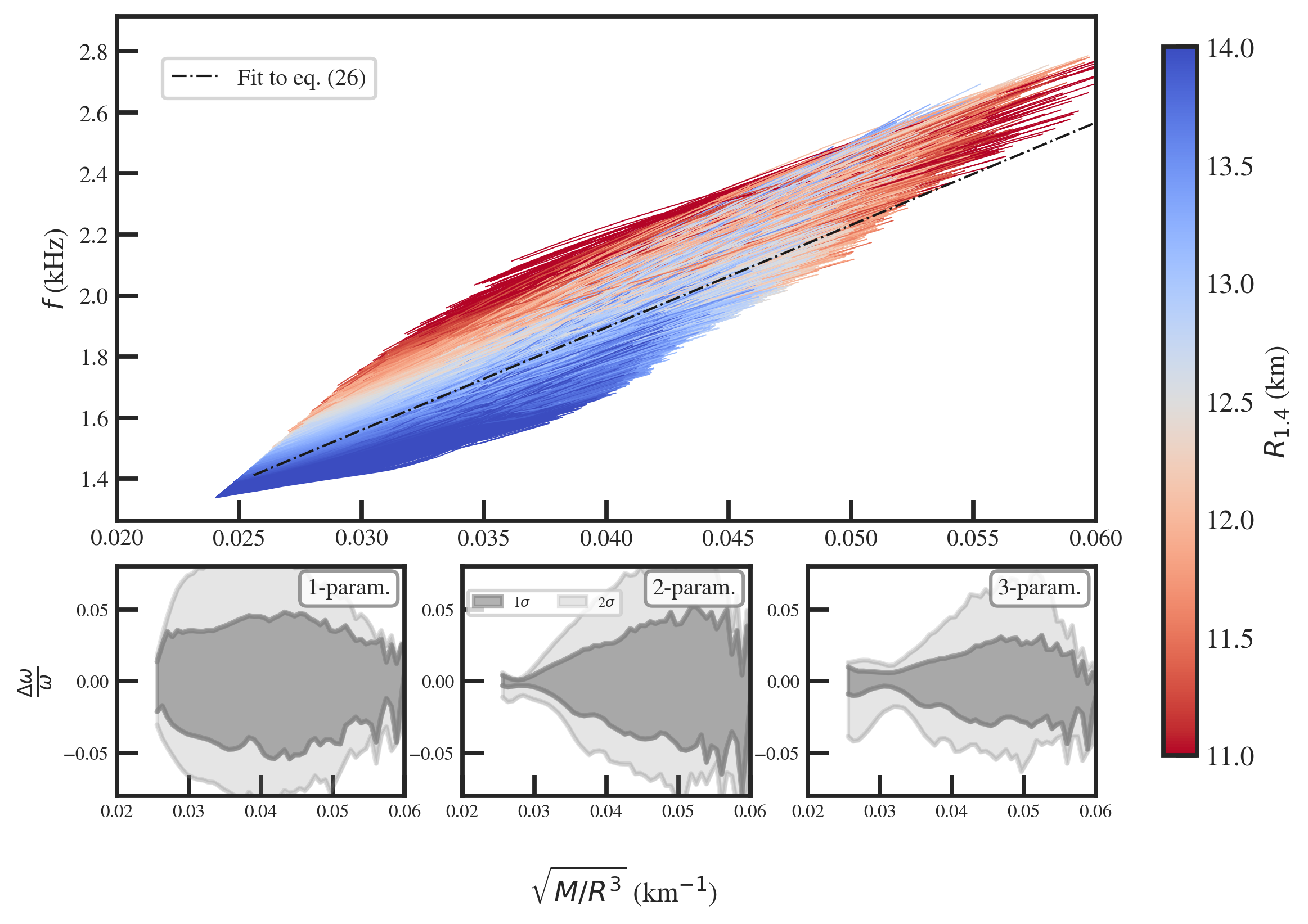}
\caption{\label{fig:avgdensity} Average density universal relation for \textit{f}-mode frequency. Each curve in the top panel represents a piecewise-polytropic EOS, with color corresponding to the canonical radius for the EOS ($R_{1.4}$). The three bottom panels are residual plots with 1$\sigma$ and 2$\sigma$ boundaries for fits to the quartic relations in average density. From left to right: the base, single-parameter relation (eq. \ref{uncorrectedquartic}); with a correction term linear in $R_{1.4}$ (eq. \ref{r1point4quartic}); and with two correction terms -- one linear in $R_{1.4}$ and one linear in $R_{1.4}/R_{1.8}$ (eq. \ref{correctedquartic}).}

\end{figure*}
\subsection{Equation of State Sample}
In order to investigate a wide portion of EOS parameter space, we construct a large sample of piecewise-polytropic EOSs, as in Refs.~\cite{Read:2008iy,Ozel:2009da,Steiner:2010fz,Raithel:2016bux}. We describe the crust at densities below the nuclear saturation density ($\rho_{\rm sat}=2.7\times10^{14}$~g/cm$^3$) with the EOS ap3 \cite{Akmal:1998cf}. At supranuclear densities, we parametrize the EOS with five polytropic segments, spaced evenly in the logarithm of density, such that the break points are placed at (1, 1.4, 2.2, 3.3, 4.9, and 7.4)$\times \rho_{\rm sat}$. The pressure along each segment is given by
\begin{equation}
P(\rho) = K_i \rho^{\Gamma_i}, \qquad \rho_{i-1} \le \rho < \rho_{i}
\end{equation}
where $\rho$ is the mass density and $K_i$ and $\Gamma_i$ are the polytropic coefficient and index, respectively, for that segment. The polytropic constant is determined by requiring continuity between adjacent segments, according to
\begin{equation}
K_i = \frac{P_{i-1}}{\rho_{i-1}^{\Gamma_{i}}} = \frac{P_i}{\rho_i^{\Gamma_i}}. 
\end{equation}
The polytropic index is then given by
\begin{equation}
\Gamma_i = \frac{\log(P_i/P_{i-1})}{\log(\rho_i/\rho_{i-1})}.
\end{equation}
We construct a sample of 30,720 EOSs by uniformly sampling the pressures at these five fiducial densities. These five pressures, along with the crust EOS below $\rho_{\rm sat}$, uniquely define the corresponding piecewise-polytropic EOS. We impose a minimal set of priors on this sample, namely that:
\begin{enumerate}
\item The EOS must be hydrostatically stable, i.e., that $\partial P/\partial \rho \ge 0$,
\item The sound speed must remain subluminal at all densities, and
\item The EOS must predict maximum masses of at least 1.97~$\Ms$, in order to be consistent with the 1$\sigma$ lower-limit on the current most massive pulsar \cite{NANOGrav:2019jur,Fonseca:2021wxt}.
\end{enumerate}

For each of the 30,720 EOSs, we compute the mass ($M$), radius ($R$), adiabatic tidal deformability ($\Lambda$), moment of inertia ($I$), \textit{f}-mode frequency ($f$ or angular frequency $\omega$), and damping time ($\tau$) using the algorithm described above. To construct our dataset, we sample each EOS at 15 evenly-spaced points in mass, from $1 \Ms$ to the maximum NS mass predicted by the respective EOS. The $M-R$, $M-\Lambda$, and $M-f$ relations for our piecewise-parametric EOS sample are shown in Fig.~\ref{fig:eossample}.

In addition, to motivate the correction terms we will introduce, Fig.~\ref{fig:R14_slope} shows the distribution of characteristic radii, $R_{1.4}$, for our sample, as well as of $R_{1.4}/R_{1.8}$, which is a parameter proportional to the slope of the mass-radius relation at intermediate masses. These parameters have been previously shown to be correlated with the pressure at 1-2$\rho_{\rm sat}$ and 3-4$\rho_{\rm sat}$, respectively \cite{Lattimer:2000nx,ozelpsaltis}. To illustrate these correlations, Fig.~\ref{fig:R14_slope} shows these parameters compared to their pressures at two fixed densities for our sample. Fig.~\ref{fig:R14_slope} shows that higher pressures at these densities tend to imply larger $R_{1.4}$ values and smaller $R_{1.4}/R_{1.8}$ values. Thus, in the following, we describe EOSs with larger $R_{1.4}$ or smaller $R_{1.4}/R_{1.8}$ slope parameters as ``stiffer", to indicate that they tend to have larger pressures at these densities.

\subsection{$\sqrt{M/R^3}-f$ Relation}
The first UR that we investigate with this sample relates the \textit{f}-mode frequency to $\sqrt{M/R^3}$, which is inversely proportional to the free-fall time for a point mass in a gravitational well, and thus sets a characteristic timescale for the oscillations. This parameter can also be regarded as the root average density. Earlier works showed that the \textit{f}-mode frequency is linearly correlated with $\sqrt{M/R^3}$ \cite{andersson,Chirenti:2015dda}, using significantly smaller (10-20) realistic, tabulated EOS samples. 
In particular, Ref.~\cite{Chirenti:2015dda} found that each EOS's \textit{f}-modes can be independently fit to a linear function of average density to $\lesssim1\%$ accuracy, but follow a combined linear fit (i.e., for the entire EOS sample) to only $\sim10\%$ accuracy, where the combined fit is of the form 
\begin{equation}\label{linearuncorrected}
    f = a_0+a_1\sqrt{\frac{M}{R^3}}
\end{equation}
and $a_n$ are fitting constants.

\begin{figure}[h]
\centering
\includegraphics[width=0.95\linewidth]{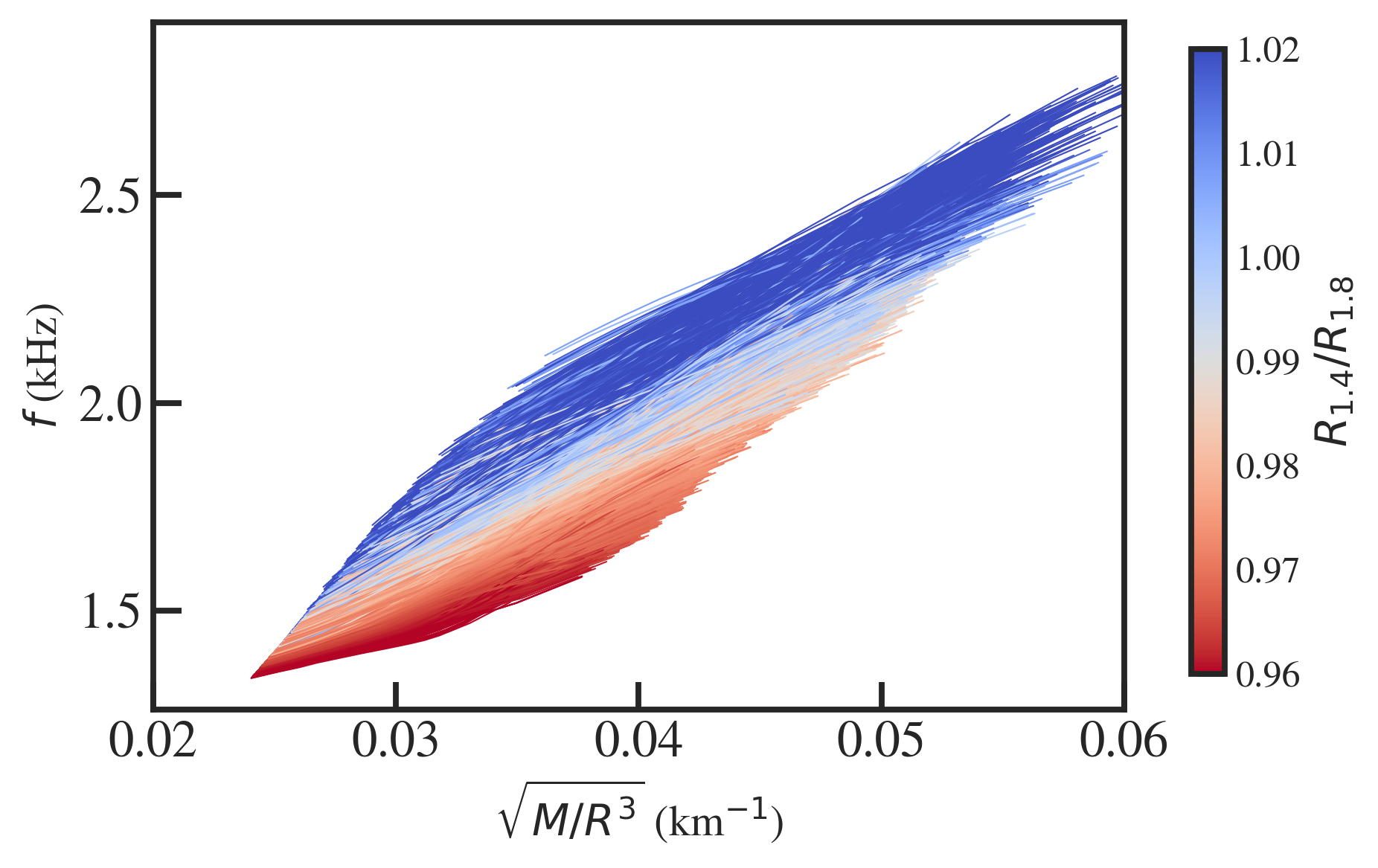}
\caption{\label{fig:mrslope} Average density universal relation for the \textit{f}-mode frequency, but color-coded with $R_{1.4}/R_{1.8}$. Each curve represents a piecewise-polytropic EOS. 
We note that a larger slope parameter (shown in blue) is generally correlated with softer EOSs, while stiffer EOSs predicting smaller slopes are shown in red (see Fig.~\ref{fig:R14_slope}).
}
\end{figure}

For each independent EOS in our large sample, we similarly find a $\lesssim1\%$ deviation for a linear fit. Across the combined sample, however, we find slightly larger residual EOS dependence in this UR, of up to $\sim$400 Hz ($\sim15-20\%$ fractional residuals), as shown in Fig.~\ref{fig:avgdensity}.  Each line in Fig.~\ref{fig:avgdensity} represents a single EOS, with the color corresponding to the predicted radius of a 1.4 $\Ms$ NS ($R_{1.4}$) for that EOS. Figure~\ref{fig:mrslope} displays the correlation with the approximate mass-radius slope $R_{1.4}/R_{1.8}$ for the same sample.

When color-coding with these parameters, our large sample exposes a clear correlation between the \textit{f}-mode frequency and the stiffness of the EOS. More specifically, the stiffest EOSs (with $R_{1.4}\gtrsim13.5$ km; shown in blue) exclusively inhabit the lower portion of Fig.~\ref{fig:avgdensity}, with $f \lesssim 1.8$ kHz at all masses. On the other hand, the softest EOSs (with $R_{1.4} \lesssim 11$ km; in red) exclusively predict $f \gtrsim 1.8$ kHz across all masses.
 
A similar trend appears for the slope parameter in Fig.~\ref{fig:mrslope}, where EOSs with $R_{1.4}/R_{1.8}\lesssim0.98$ (in red) have $f<2$ kHz at all masses, and EOSs with slope parameters $\gtrsim 1$ (in blue) generally have larger $f$, though spanning a larger range.

We quantify these correlations by performing a linear regression between the residuals from our fit to the average density UR (in the form of eq.~\ref{uncorrectedquartic}) and the parameters $R_{1.4}$ and $R_{1.4}/R_{1.8}$. In particular, we calculate the $R^2$ coefficients of determination for the residuals at an average density value of 0.036 km$^{-1}$, which corresponds to a $1.4~\Ms$ NS with the SLy4 EOS \cite{Douchin:2001sv}, which predicts a radius of 11.7~km. At this characteristic value, we find strong correlations between the residuals to the UR and our two EOS-sensitive parameters, with $R^2$ values greater than 0.7. We report the correlation coefficients in Table \ref{tab:avgdrsquared}.
 
\begin{table}[h]
\centering
\begin{tabular}{c|c|c}
    \multicolumn{3}{c}{\textbf{$\sqrt{M/R^3}-f$}} \\ \noalign{\hrule height 1.3pt}
     \# Params. & $R^2$ with $R_{1.4}$ & $R^2$ with $\frac{R_{1.4}}{R_{1.8}}$ \\ \noalign{\hrule height 1.3pt}
     1 & 0.769 & 0.713 \\
    2 & 0.0336 & 0.445 \\
    3 & 0.0748 & 0.201 \\
\end{tabular}
\caption{The coefficient of determination between $R_{1.4}$ or slope and the residuals for the average-density relation, evaluated at $\sqrt{M/R^3}=0.0357$ km$^{-1}$ (corresponding to a $1.4 \Ms$ SLy4 NS). We find significant correlations in the residuals with both $R_{1.4}$ and the slope, which are effectively removed with the 3-parameter UR.}
\label{tab:avgdrsquared}
\end{table}
 
These correlations reveal a residual dependence in the single-parameter UR on the stiffness of the EOS. We seek to eliminate this dependence by introducing linear correction terms to the UR that depend on $R_{1.4}$ and the slope. 
First, to increase the overall accuracy of the average density fit, we follow Ref.~\cite{Chan:2014kua}'s expansion of the \textit{f}-Love relation to fourth-order, taking the single-parameter $\sqrt{M/R^3}$ fit to fourth-order as well. We also add linear correction terms to create 2- and 3-parameter URs, so that we consider the following 3 models for the \textit{f}-mode frequency: 
\begin{subequations}
\begin{equation}\label{uncorrectedquartic}
    f = a_0 + a_1x+ a_2x^2 + a_3x^3+ a_4x^4 
\end{equation}
\begin{equation}\label{r1point4quartic}
        f = a_0 + a_1x+ \cdots+ a_4x^4 + b_1R_{1.4} 
\end{equation}
\begin{equation}\label{correctedquartic}
        f = a_0 + a_1x+ \cdots+ a_4x^4 + b_1R_{1.4} + c_1\Bigg(\frac{R_{1.4}}{R_{1.8}}\Bigg)
\end{equation}
\end{subequations}where $x=\sqrt{\frac{M}{R^3}}$ and $a_n$, $b_n$, and $c_n$ are fit parameters.
Table \ref{tab:fresid} reports the average 1$\sigma$ residuals for both the entire sample (with every EOS sampled with 15 points spaced uniformly in mass from 1~$\Ms$ to $M_{\rm max}$) and also for a sub-sample of one $1.4\Ms$ NS sampled from every EOS. To calculate the 1$\sigma$ value, we take the standard deviation of the full set of residuals for each respective sample. Table \ref{tab:coeffs} in Appendix~\ref{sec:appendixparams} reports our best-fit coefficients to each of these URs. We find that the 3-parameter model significantly improves the fit, bringing the full-sample 1-$\sigma$ residual from 3.8\% down to 1.5\% (a 60\% fractional change) and for 1.4$\Ms$ NSs, bringing the 1-$\sigma$ residuals from 3.3\% down to 0.8\% (a 75\% fractional change).

We show the 1- and 2-$\sigma$ residuals as a function of average density in the bottom row of Fig.~\ref{fig:avgdensity}, with the 1-, 2-, and 3-parameter fits from left to right. Considering the 1-$\sigma$ band in particular, we see that the 2-parameter fit exhibits a marked improvement in the residuals at low average densities compared to the 1-parameter fit, bringing the 1-$\sigma$ errors in this regime from $\sim$4\% to $<1\%$, but that the residuals get slightly worse at high average densities. When the slope correction term is added as well (shown on the right), there is a large improvement at high average densities, taking the 1-$\sigma$ errors down to $\sim$2\%, while roughly maintaining the accuracy at low average densities. 

For these 2- and 3-parameter fits, we observe smaller correlations between the residuals to these expanded URs and either $R_{1.4}$ or the slope (see Table~\ref{tab:avgdrsquared}). For the 3-parameter version of the UR, we find that the final residuals are minimally correlated with $R_{1.4}$ or the slope, with $R^2$ values of 0.075 and 0.20, respectively, confirming that we have effectively removed the $R_{1.4}$-dependence and most of the slope-dependence from this UR. 

\begin{table}[h]
\centering
\begin{tabular}{!{\vrule width 1.5pt}c|c|c|c|c|c!{\vrule width 1.5pt}}
    \noalign{\hrule height 1.5pt}
     & \# Params. & Avg. 1-$\sigma$ & \% Red. & $1.4\Ms$ 1-$\sigma$ & \% Red. \\ \noalign{\hrule height 1.5pt}
    \multirow{3}{1.5cm}{$\sqrt{\frac{M}{R^3}}-f$} & 1 & 3.76$\times10^{-2}$ & --- & 3.33$\times10^{-2}$ & --- \\ \cline{2-6} 
        & 2 & 1.74$\times10^{-2}$ & 53.6 & 4.70$\times10^{-3}$ & 85.9 \\ \cline{2-6} 
       & 3 & 1.50$\times10^{-2}$ & 60.0 & 8.46$\times10^{-3}$ & 74.6 \\ \noalign{\hrule height 1.5pt}
    \multirow{3}{1.5cm}{$\eta-M\omega$} & 1 & 7.91$\times10^{-4}$ & --- & 8.07$\times10^{-4}$ & --- \\ \cline{2-6} 
     & 2 & 7.06$\times10^{-4}$ & 10.8 & 6.47$\times10^{-4}$ & 19.9 \\ \cline{2-6} 
     & 3 & 6.36$\times10^{-4}$ & 19.6 & 5.40$\times10^{-4}$ & 33.1 \\ \noalign{\hrule height 1.5pt}
    \multirow{3}{1.5cm}{\textit{f}-Love} & 1 & 3.66$\times10^{-3}$ & --- & 3.84$\times10^{-3}$ & --- \\ \cline{2-6} 
     & 2 & 3.55$\times10^{-3}$ & 3.1 & 3.61$\times10^{-3}$ & 5.9 \\ \cline{2-6} 
     & 3 & 3.46$\times10^{-3}$ & 5.4 & 3.47$\times10^{-3}$ & 9.5 \\ \noalign{\hrule height 1.5pt}
\end{tabular}
\caption{Residuals for \textit{f}-mode frequency URs. We report the average $1\sigma$ residuals for the full-sample (i.e., with 15 points sampled for each EOS, uniformly between 1$\Ms$ and $M_{\rm max}$), as well as the average 1$\sigma$ residuals for the sub-sample of 1.4$\Ms$ NSs. For each of these measures, we also calculate the fractional reduction in residuals with the multi-parameter fits, compared to the residuals for the single-parameter fit, and report these percent reductions in the columns labeled ``\% Red".}
\label{tab:fresid}
\end{table}
\begin{figure*}[ht]
\centering
\includegraphics[width=0.85\linewidth]{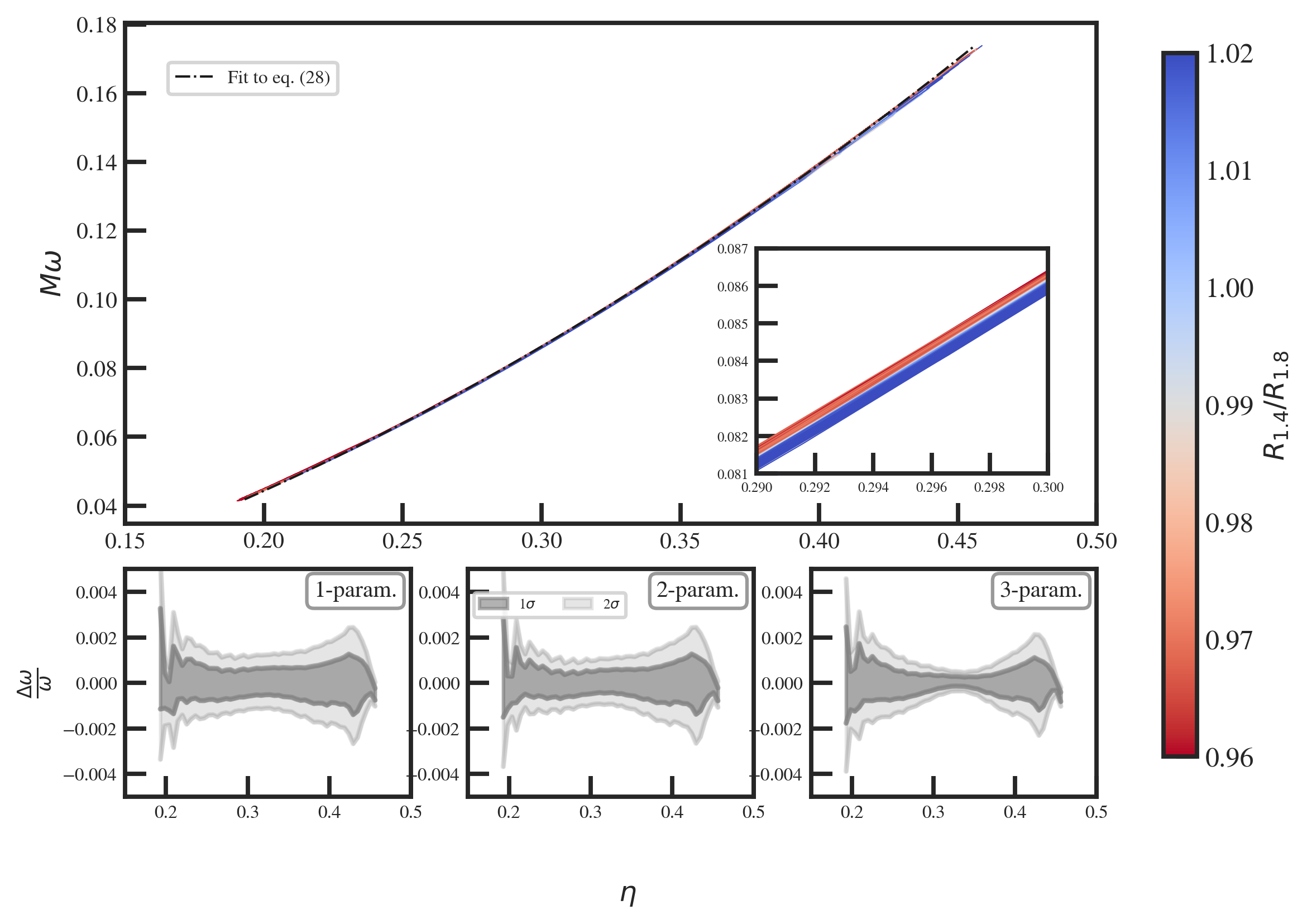}
\caption{\label{fig:etamomega} Universal relation between $M\omega$ and the effective compactness, $\eta$. Each curve in the top panel represents a piecewise-parametric EOS -- note that all 30,720 EOSs initially seem to form one single line. However, the inset shows a smaller region about the $1.4\Ms$ SLy4 value ($\eta =0.294$), where a trend with $R_{1.4}/R_{1.8}$ is evident ($R^2=0.7)$. The three bottom panels are residual plots with 1$\sigma$ and 2$\sigma$ boundaries for fits to the quintic relations in $\eta$. From left to right: the base (single-parameter) relation, with a linear correction term in $R_{1.4}$, and with two correction terms -- one linear in $R_{1.4}$ and one linear in $R_{1.4}/R_{1.8}$.}
\end{figure*}

\subsection{$\eta -M\omega$ Relation}
The next UR we consider relates the $f$-mode angular frequency, $\omega$, to a dimensionless effective compactness $\eta \equiv \sqrt{M^3/I}$, as was first introduced in Ref.~\cite{Lau:2009bu}. This was used to construct the $\eta - M\omega$ relation
\begin{equation}
\label{eq:eta}
    M\omega = a_0 + a_1\eta + a_2 \eta^2,
\end{equation} which yielded errors of 0.37\% and 0.7\%, respectively, for their original sample of 9 NS EOSs and 2 quark star EOSs \cite{Lau:2009bu}.

As for the average density UR, we initially expanded eq.~(\ref{eq:eta}) to quartic order in $\eta$. However, there was a clear trend in the shape of the residuals, which was eliminated by taking the fit to quintic order in $\eta$. Thus, we adopt a quintic base UR for the rest of this section. Our larger EOS sample shows that this relation has far less scatter (i.e., less EOS dependence) than the average density relation to begin with, with a full-sample 1-$\sigma$ residual and $1.4\Ms$ 1-$\sigma$ residual of 0.079$\%$ and 0.081$\%$, respectively, along with $2\sigma$ residuals that are kept at or below $0.2\%$ for the single-parameter UR (see Fig.~\ref{fig:etamomega} and Table \ref{tab:fresid}).

To search for residual stiffness correlations within this relation, the inset in Fig.~\ref{fig:etamomega} zooms in around the region $0.29<\eta<0.3$, corresponding to the characteristic value for a 1.4~$\Ms$ NS with the SLy4 EOS ($\eta=0.294$). We do not find a strong correlation with $R_{1.4}$; however the inset in Fig.~\ref{fig:etamomega} reveals a clear trend with $R_{1.4}/R_{1.8}$. Indeed, although the residuals to the single-parameter fit are small, they have a significant correlation with the slope of $R^2$=0.701, as reported in Table~\ref{tab:etafrsquared}. Accordingly, after applying linear correction terms in $R_{1.4}$ and slope (in the same way eqs.~\ref{r1point4quartic} and \ref{correctedquartic} build upon eq.~\ref{uncorrectedquartic}), there is a reduction in the full-sample and $1.4\Ms$ 1-$\sigma$ residuals of $\sim20\%$ and $\sim 33\%$ of their respective single-parameter magnitudes. In other words, we find that, although the $\eta-M\omega$ UR is already very tight (sub-percent residuals), the remaining EOS-dependence is highly correlated with the slope and, by removing this trend, we can further reduce the $1.4\Ms$ 1-$\sigma$ residuals by an additional 33\%. Qualitatively, we observe the residuals for the 1- and 2-parameter fits are approximately uniform for all values of $\eta$, although the shape tightens significantly at intermediate $\eta$ with the addition of the $R_{1.4}/R_{1.8}$ term (see the bottom row of Fig.~\ref{fig:etamomega}). We confirm that we successfully eliminate the correlation with slope in the residuals of the 3-parameter fit, as evidenced by a final $R^2$ value of 0.10 (see Table~\ref{tab:etafrsquared}). 

\begin{table}[h]
\centering
\begin{tabular}{c|c|c}
    \multicolumn{3}{c}{\textbf{$\eta - M\omega$}} \\ \noalign{\hrule height 1.3pt}
     \# Params. & $R^2$ with $R_{1.4}$ & $R^2$ with $\frac{R_{1.4}}{R_{1.8}}$ \\ \noalign{\hrule height 1.3pt}
    1 & 0.291 & 0.701 \\
    2 & 0.00119 & 0.298 \\
    3 & 0.000231 & 0.102 \\
\end{tabular}
\caption{The coefficient of determination between $R_{1.4}$ or slope and the residuals for the effective compactness relation at $\eta = 0.294$, corresponding to a $1.4 \Ms$ SLy4 NS. When comparing the single-parameter residuals to those of the average density relation, there is clearly less correlation with $R_{1.4}$, but still a comparable degree of correlation with slope. }
\label{tab:etafrsquared}
\end{table}

\subsection{\textit{f}-Love Relation}
The final UR we consider for the $f$-mode frequency relates to the adiabatic tidal deformability. This $f$-Love relation was first introduced in Ref.~\cite{Chan:2014kua} and can be viewed as a corollary of eq.~(\ref{eq:eta}), through the I-Love-Q relation, which closely relates a NS's moment of inertia and adiabatic tidal deformability  \cite{iloveq}. The \textit{f}-Love relation is of particular interest because of its applications to GWs and NS mergers \cite[e.g.,][]{Pratten:2021pro,schmidthinderer,williams,Pradhan:2022rxs}. The proposed relation from Ref.~\cite{Chan:2014kua} has the form 
\begin{equation}
    M \omega = a_0 + a_1x + a_2x^2 + a_3x^3 + a_4 x^4,
    \label{eq:fLove}
\end{equation}
where $x = \ln \Lambda$. This UR has little EOS-sensitivity for the small sample of polytropic and tabulated EOSs that is presented in Ref.~\cite{Chan:2014kua}, with residuals below 1\% for all but the most extreme masses. 

\begin{figure}[ht]
\centering
\includegraphics[width=0.95\linewidth]{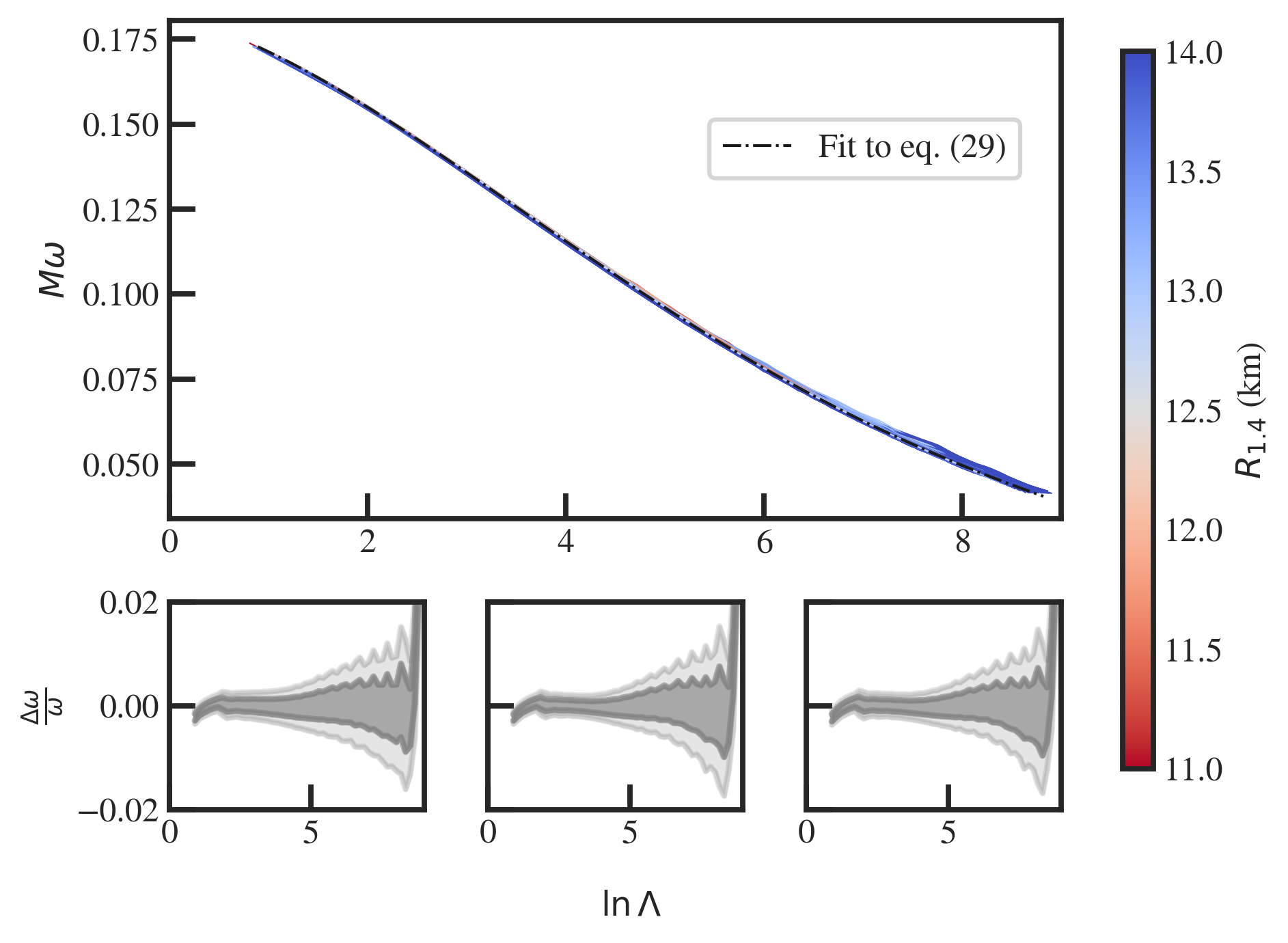}
\caption{\label{fig:flove} \textit{f}-Love universal relation for the scaled \textit{f}-mode $M\omega$. Each curve in the top panel represents a piecewise-parametric EOS -- note that all 30,720 EOSs fall nearly on top of one another. The three bottom panels are residual plots with 1$\sigma$ and 2$\sigma$ boundaries for the quartic relation in $\ln \Lambda$. From left to right: the base (single-parameter) relation, with a linear correction term in $R_{1.4}$, and with two correction terms -- one linear in $R_{1.4}$ and one linear in $R_{1.4}/R_{1.8}$.} 
\end{figure}

Using our larger EOS sample, we observe comparably little EOS-dependence in this UR, with even the $2\sigma$ residuals falling within $\pm$1\% bounds for all but the largest values of $\ln \Lambda$, as shown in Fig.~\ref{fig:flove}. For large values of $\Lambda$, corresponding to the least compact stars, there is more EOS-sensitivity, but the $2\sigma$ residuals still remain within $\pm2\%$ up until the very boundary of the sample.
Additionally, trends with $R_{1.4}$ or $R_{1.4}/R_{1.8}$ in the deviation from universality are minimal, with $R^2$ values of only 0.240 and 0.0241, respectively (see Fig.~\ref{fig:flove} and Table \ref{tab:floversquared}). Thus, we find that the $f$-Love UR has no significant residual correlation with these EOS stiffness parameters.

\begin{table}[h]
\centering
\begin{tabular}{c|c|c}
\multicolumn{3}{c}{$f$-Love} \\ \noalign{\hrule height 1.3pt}
     \# Params. & $R^2$ with $R_{1.4}$ & $R^2$ with $\frac{R_{1.4}}{R_{1.8}}$ \\ \noalign{\hrule height 1.3pt}
    1 & 0.240 & 0.0241 \\
    2 & 0.0144 & 0.0134 \\
    3 & 0.0152 & 0.000643 \\
\end{tabular}
\caption{The coefficient of determination between $R_{1.4}$ or slope and the residuals for the \textit{f}-Love relation at $\ln \Lambda = 5.72$, corresponding to a $1.4 \Ms$ SLy4 NS. There is minimal correlation with $R_{1.4}$ and effectively no correlation with slope. }
\label{tab:floversquared}
\end{table}

For completeness, we generate 2- and 3-parameter fits that are built upon the quartic single-parameter relation, adding linear corrections in $R_{1.4}$ and $R_{1.4}/R_{1.8}$ following eqs. (\ref{r1point4quartic}) and (\ref{correctedquartic}) with $x=\ln\Lambda$. As expected, based on the lack of $R_{1.4}$ or slope correlation in the single-parameter UR residuals, the \textit{f}-Love UR displays no qualitative change in the shape of the residuals for the higher-parameter fits, as shown in the bottom row of Fig.~\ref{fig:flove}. The magnitude of the residuals remains about constant as well, with the 3-parameter relation seeing fractional reductions of $5.4\%$ and $9.5\%$ in the full-sample and $1.4\Ms$ 1-$\sigma$ values, respectively (see Table \ref{tab:fresid}). Although the $R^2$ values are reduced by at least an order of magnitude with the inclusion of the $R_{1.4}$ and slope terms, we note that the 1-parameter $f$-Love relation is already highly EOS-insensitive and lacking residual stiffness correlations.

\subsection{$\tau$ Relations}
\label{sec:damping}
There also exist URs for the damping time of \textit{f}-mode oscillations, which is the inverse of the imaginary component of the \textit{f}-mode eigenfrequency. Similar to the URs for the real part of the \textit{f}-mode, these have been posed in terms of a combination of mass and radius, in terms of $\eta$, and in terms of $\Lambda$ \cite{Tsui:2004qd,Lau:2009bu,PhysRevD.104.123002,andersson,Sotani:2021kiw,Chirenti:2015dda}.
\begin{figure}
    \centering
    \includegraphics[width=0.95\linewidth]{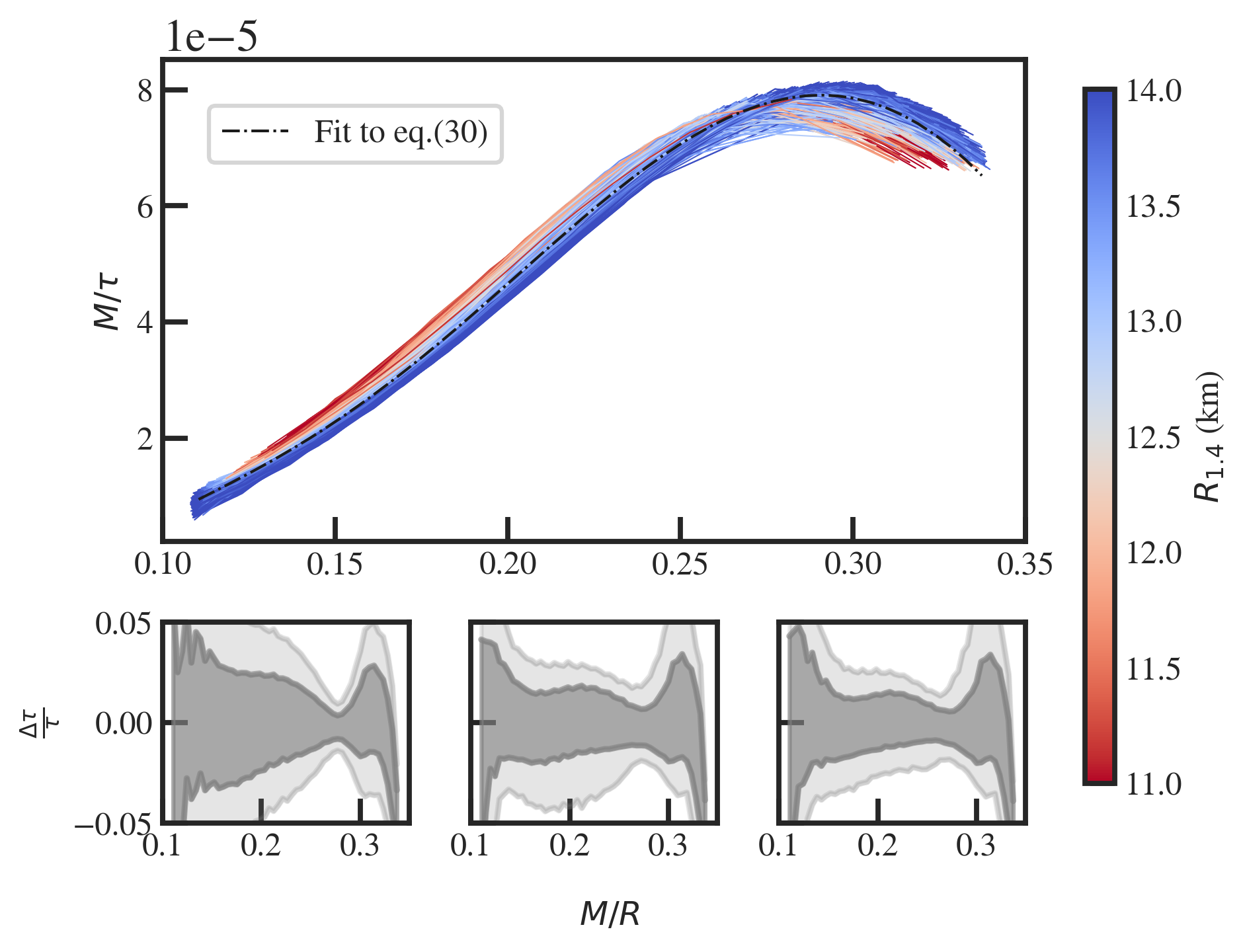}
    \caption{UR between compactness and $M/\tau$. The 3 subplots at the bottom show the 1-$\sigma$ and $2\sigma$ residuals for fits that are quintic in $M/R$. From left to right: the base (single-parameter) relation, with a linear correction term in $R_{1.4}$, and with linear correction terms in both $R_{1.4}$ and the slope.}
    \label{fig:mtaucompact}
\end{figure}

We explore existing URs which are  written in terms of the mass-scaled imaginary component of the eigenfrequency, i.e. $M/\tau$. In particular, we consider models of the form
\begin{equation} \label{dimfullbase}
    M/\tau = a_0 + a_1x+a_2x^2+a_3x^3+a_4x^4+a_5x^5,
\end{equation}
where $x$ is either compactness ($M/R$) or $\Lambda^{-1/5}$, which we will refer to as the reduced tidal deformability. We elect to expand to quintic order in order to eliminate a clear trend in the residuals for the reduced tidal deformability fit. 

One early version of this relation was introduced in Ref.~\cite{Lau:2009bu}, where $M/\tau$ was related to the compactness, at quadratic order. With a sample of 9 tabulated NS EOSs and 2 quark star EOSs, the quadratic fit that they employed captured the general trend of the compactness$-M/\tau$ relation, but did not fully capture the distribution of damping times, resulting in errors of $\sim50\%$ for the quark stars and $>200\%$ for the NSs \cite{Lau:2009bu}. This same work also tested an $\eta-M/\tau$ relation and found that, although the $M/\tau$ values all fell on a common smooth curve, a quadratic fit was insufficient to capture the trend with $\eta$. A tighter UR was found in Ref.~\cite{PhysRevD.104.123002}, by going to quintic-order in the fit and by comparing $M/\tau$ to $\Lambda$. This yielded residuals of $\lesssim1\%$ for a sample of 8 tabulated EOSs.

Our choice of $\Lambda^{-1/5}$ is motivated by its scaling in the Newtonian limit for an $n=0$ or 1 polytrope, for which $\Lambda \propto (M/R)^{-5}$. In the same limit, $\eta \propto(M/R)$, and thus $\eta \propto \Lambda^{-1/5}$ \cite{iloveq}. So, we expand upon the briefly-explored $\eta-M/\tau$ relation presented in Ref.~\cite{Lau:2009bu} by replacing $\eta$ with the reduced tidal deformability $\Lambda^{-1/5}$, to investigate whether this better captures the universality. We note that this parameter has been previously used in UR fits for the $f$-mode frequency by Ref.~\cite{PhysRevD.104.043011}, where they found comparable universality for fits with $\ln\Lambda$ and $\Lambda^{-1/5}$.

\begin{table}[h]
\centering
\begin{tabular}{c|c|c!{\vrule width 1.5pt}c|c|c}
    \multicolumn{3}{c}{\textbf{$M/R - M/\tau$}} & \multicolumn{3}{c}{\textbf{$\Lambda^{-1/5} - M/\tau$}} \\ \noalign{\hrule height 1.3pt}
     Pars. &  $R^2$ w/ $R_{1.4}$ & $R^2$ w/ $\frac{R_{1.4}}{R_{1.8}}$ & Pars. & $R^2$ w/ $R_{1.4}$  & $R^2$ w/ $\frac{R_{1.4}}{R_{1.8}}$ \\ \noalign{\hrule height 1.3pt}
    1 & 0.679 & 0.699 & 1 & 0.116 & 0.104 \\
    2 & 0.268 & 0.618 & 2 & 0.0731 & 0.00663 \\
    3 & 0.411 & 0.213 & 3 & 0.0713 & 0.0659 \\
\end{tabular}
\caption{The coefficient of determination between $R_{1.4}$ or slope and the residuals for the $M/\tau$ fits, at characteristic values corresponding to a $1.4 \Ms$ SLy4 NS. }
\label{tab:basetaursquared}
\end{table} 

\begin{figure}
    \centering
    \includegraphics[width=0.95\linewidth]{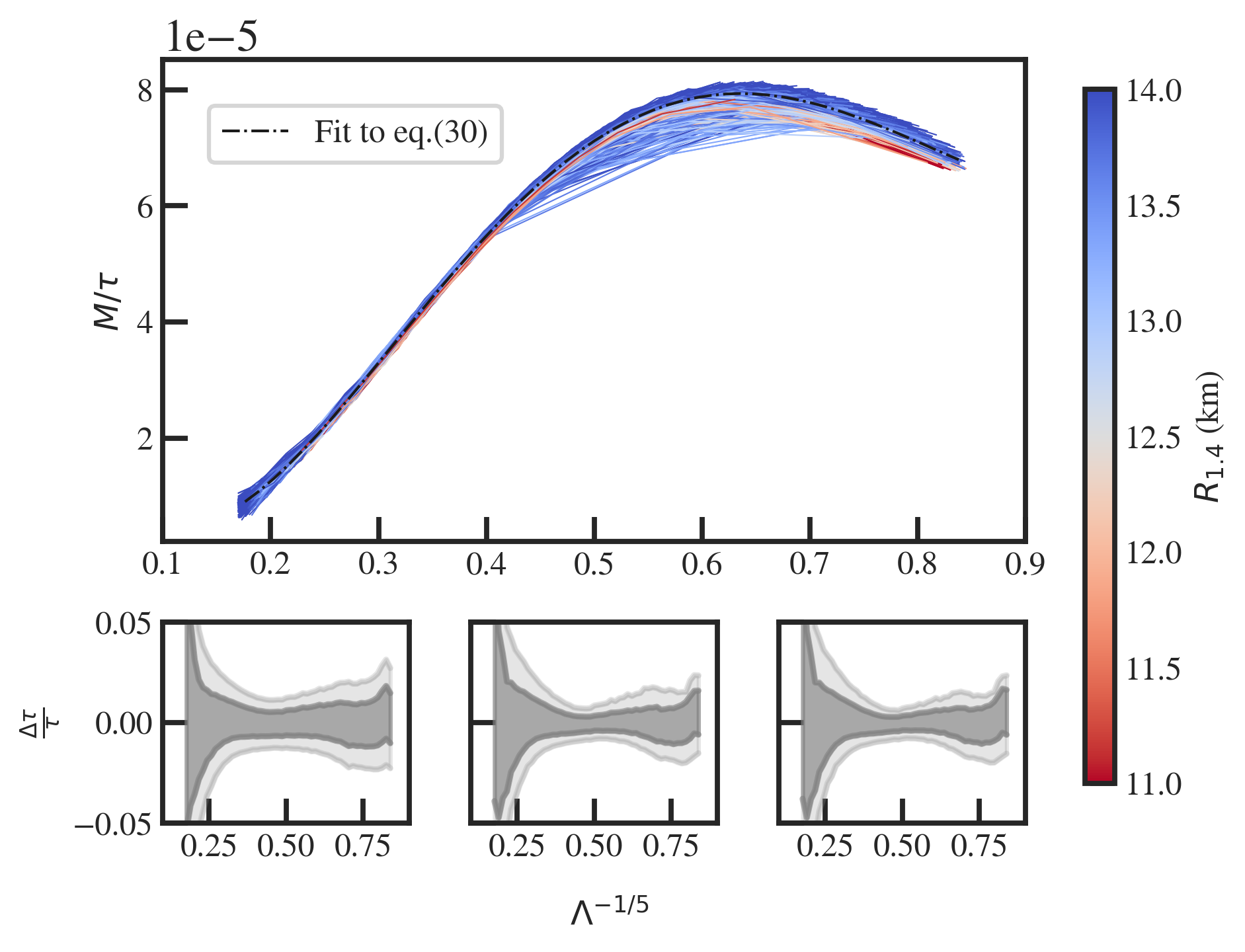}
    \caption{UR between the reduced tidal deformability and $M/\tau$. The 3 subplots at the bottom show the 1-$\sigma$ and $2\sigma$ residuals for quintic fits in $\Lambda^{-1/5}$. From left to right: the base (single-parameter) relation, with a linear correction term in $R_{1.4}$, and with linear correction terms in both $R_{1.4}$ and the slope.}
    \label{fig:mtaulambda}
\end{figure}

We expand on these previous results by fitting eq.~(\ref{dimfullbase}) with our parametric sample of 30,720 EOSs. We show the results for the fit with $x=M/R$ in Fig.~\ref{fig:mtaucompact} and with $x=\Lambda^{-1/5}$ in Fig.~\ref{fig:mtaulambda}. For the compactness fit in Fig.~\ref{fig:mtaucompact}, we find 1-$\sigma$ residuals at $\lesssim5\%$ at all masses (see the bottom left panel of Fig.~\ref{fig:mtaucompact}). The residuals are largest for very high or very low compactness values, with a strong tightening of the relation around $M/R=0.28$. We find significant correlations in the residual EOS-sensitivity for this UR with both $R_{1.4}$ and the slope, with $R^2\sim0.7$ (see Table \ref{tab:basetaursquared}). We perform additional fits with linear correction terms in $R_{1.4}$ and $R_{1.4}/R_{1.8}$ added to eq.~(\ref{dimfullbase}) for this relation. This leads to significant improvement in the UR between $0.15\lesssim M/R\lesssim 0.27$, as shown in the bottom panels of Fig.~\ref{fig:mtaucompact}. Quantitatively, we find that the compactness fit experiences $\sim20\%$ fractional reduction in the full-sample 1-$\sigma$ residual and $\sim45\%$ fractional reduction in the $1.4\Ms$ sub-sample 1-$\sigma$ residual with the addition of the two stiffness correction terms, as reported in  Table~\ref{tab:tauresid}.

\begin{table}[h]
\centering
\begin{tabular}{!{\vrule width 1.5pt}c|c|c|c|c|c!{\vrule width 1.5pt}}
    \noalign{\hrule height 1.5pt}
     & \# Params. & Avg. 1-$\sigma$ & \% Red. & $1.4\Ms$ 1-$\sigma$ & \% Red. \\ \noalign{\hrule height 1.5pt}
    \multirow{3}{1.5cm}{$M/R-M/\tau$} & 1 & 2.82$\times 10^{-2}$ & --- & 2.98$\times 10^{-2}$ & --- \\ \cline{2-6} 
        & 2 & 2.23$\times 10^{-2}$ & 20.8 & 1.79$\times 10^{-2}$ & 39.9 \\ \cline{2-6} 
       & 3 & 2.31$\times 10^{-2}$ & 18.0 & 1.63$\times 10^{-2}$ & 45.2 \\ \noalign{\hrule height 1.5pt}
    \multirow{3}{1.5cm}{$\Lambda^{-1/5} - M/\tau$} & 1 & 1.81$\times 10^{-2}$ & --- & 1.38$\times 10^{-2}$ & --- \\ \cline{2-6} 
     & 2 & 1.86$\times 10^{-2}$ & -2.87 & 1.37$\times 10^{-2}$ & 0.753 \\ \cline{2-6} 
     & 3 & 1.87$\times 10^{-2}$ & -3.09 & 1.35$\times 10^{-2}$ & 1.77 \\ \noalign{\hrule height 1.5pt}
\end{tabular}
\caption{Residuals for the mass-scaled damping time URs. We report the average 1-$\sigma$ residuals for the full-sample and for the sub-sample of 1.4~$\Ms$ NSs, along with the percent reductions in magnitude for residuals with the multi-parameter fits compared to the single-parameter fits. Note that a negative percentage reduction implies an increase in the value of the residual compared to the single-parameter value.}
\label{tab:tauresid}
\end{table}

In contrast, Fig.~\ref{fig:mtaulambda} shows that the analogous baseline fit for tidal deformability  (eq.~\ref{dimfullbase} with $x=\Lambda^{-1/5}$) yields 1-$\sigma$ residuals that are significantly smaller, between $\sim0.5-2\%$ for the majority of the interval. We find negligible correlations between these residuals and either $R_{1.4}$ or the slope. For completeness, we perform additional fits with linear correction terms in $R_{1.4}$ and $R_{1.4}/R_{1.8}$ for this UR as well. However, we find these corrections yield only minor reductions in the $1.4\Ms$ sub-sample residual, and actually have small increases in the full-sample average (see Table \ref{tab:tauresid}), suggesting that these correction terms are not justified to include for this relation.

Finally, we note that we also explored a version of eq.~(\ref{dimfullbase}) with $x=\eta$ and find that
the $\eta-M/\tau$ relation performs nearly identically to the $\Lambda^{-1/5}$ relation, in both the baseline residuals and the response to the stiffness parameters. This is perhaps not surprising, given the strong I-Love universality between the two parameters \cite{iloveq}. 

Overall, the strength of the correlations in the UR residuals with the EOS stiffness parameters follows the same pattern for the damping time relations as we found for the frequency relations: i.e., URs in terms of combinations of mass and radius have the largest residuals and the strongest correlation with $R_{1.4}$ or slope, while URs in terms of the effective compactness $\eta$ or tidal deformability have smaller residuals and less correlation with the stiffness parameters.

There are URs for other dimensionless versions of the damping time that have been explored as well \cite[e.g.,][]{andersson,Lau:2009bu,Sotani:2021kiw}, which we find perform similarly to those shown in Figs.~\ref{fig:mtaucompact} and \ref{fig:mtaulambda}. We include several of these in Appendix~\ref{sec:appendixdamp} for completeness.

\section{Application to Pulsar Glitch Observables} \label{sec:glitches}
To demonstrate how residual stiffness correlations in a UR can affect predicted observables, we explore their effect on the amplitude of GWs from a Vela-like glitching pulsar. Pulsar glitches occur when a rapidly rotating NS (a pulsar) experiences a sudden increase in its decaying spin frequency, as a result of friction between the crust and the superfluid beneath it \cite{Antonopoulou:2022rpq}. If this disturbance were to excite \textit{f}-mode oscillations of the pulsar, this process could generate observable gravitational radiation. 

Measuring these gravitational wave signals could be another means of learning about NS structure, e.g., through Bayesian inferences \cite{Pradhan:2023zor} or from differences in the distributions of \textit{f}-modes for quark star (QS) and NS EOSs, with QS EOSs having a narrower predicted distribution than NS EOSs \cite{ho_followup}. In both of these examples \cite{Pradhan:2023zor, ho_followup}, URs were used to predict the $f$-mode frequencies, with $f$-Love used for the former and the average-density UR (eq.~\ref{linearuncorrected}, in particular) for the latter.

To date, there have been no confirmed detections of $f$-mode GWs from a pulsar glitch. Recent analysis from the LVK collaboration on O4 data has placed upper limits on the energies of a Vela pulsar glitch diverted to the $f$-mode based on either a short-burst or continuous wave model \cite{Abac_2026}. Although these upper limits are starting to be constraining at sub-kilohertz frequencies, no meaningful upper bounds were set for the $>1$ kHz regime, which is where $f$-mode signals are almost certain to reside. Thus, we assume that the entirety of the energy of the glitch is distributed to the pulsar's $f$-mode in the analysis that follows.

We follow Ref.~\cite{ho_primary}'s treatment of the signal-to-noise ratio (SNR) of these events:
\begin{equation} \label{eq:glitchSNR}
\begin{split}
    \text{SNR} = 2&.28 \times 10^{-24} \Big(\frac{\text{1 kpc}}{d}\Big) \Big( \frac{I}{10^{45}\text{ g cm}^2} \Big)^{1/2}\Big( \frac{\nu_s}{\text{10 Hz}} \Big)^{1/2}  \\
    \times&\Big( \frac{\Delta\nu_s}{\text{$10^{-7}$ Hz}} \Big)^{1/2}\Big(\frac{\text{1 kHz}}{\nu_{gw}}\Big)\Big(\frac{1}{2S_h}\Big)^{1/2},
\end{split}
\end{equation}
where $d$ is distance to the pulsar, $\nu_s$ is the spin frequency of the pulsar, $\Delta\nu_s$ is the size of the glitch, $\nu_{gw}$ is the \textit{f}-mode frequency, and $S_h$ is the sensitivity of the detector at that particular frequency. While Ref.~\cite{ho_primary} assumes that $I \sim 10^{45}$ g cm$^2$ for all NSs, we use the actual moment of inertia vs. mass relations from our sample. 

To simulate the actual population of pulsar masses, we assume a Gaussian NS mass distribution about $1.4 \Ms$ with a standard deviation of $0.15 \Ms$ and draw 15 NSs for each of the piecewise-parametric EOSs \cite{ho_primary}. We use this synthetic dataset to obtain predictions for the SNR in the ALIGO detector of a 20 $\mu$Hz glitch of the Vela pulsar, which has a well-documented distance and spin rate \cite{vela_dist,vela_glitch}.

\begin{figure*}[t]
\centering
\includegraphics[width=0.95\textwidth]{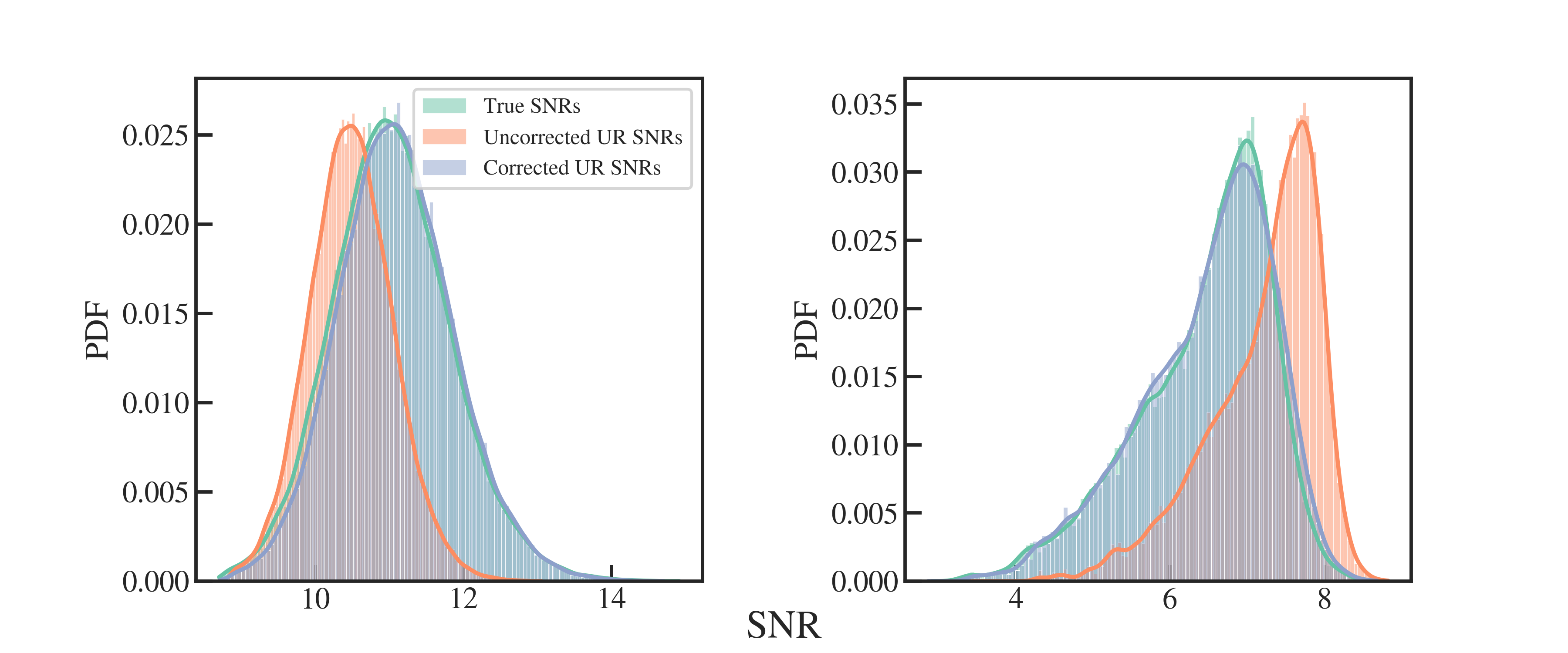}
\caption{\label{fig:SNRfig} Distribution of predicted SNR values for a GW event triggered by a 20 $\mu$Hz Vela pulsar glitch, as observed with the sensitivity of ALIGO, following eq.~(\ref{eq:glitchSNR}). We assume a Gaussian mass distribution about 1.4 $M_\odot$ with a standard deviation of 0.15 $M_\odot$. The left panel displays the stiffest EOSs in our sample (i.e., those with $R_{1.4}$ 1$\sigma$ above the mean of our sample). The right panel displays the soft EOSs (i.e., those with $R_{1.4}$ 1$\sigma$ below the mean of our sample).}
\end{figure*}

The final piece needed to estimate the SNR for this population is the $f$-mode frequency, $\nu_{\rm gw}$. For a pulsar of a given mass in our distribution, we estimate $\nu_{\rm gw}(M)$ in one of three ways: (1) using the full-GR $f$-mode calculation, described in Sec.~\ref{sec:methods}; (2) using the single-parameter average-density UR from eq. (\ref{linearuncorrected}); or (3) using the 3-parameter average-density UR from eq. (\ref{correctedquartic}), which contains both stiffness correction terms.
To emphasize the impact of the residual stiffness-correlations, we consider the stiffest and softest EOSs separately: i.e., those with $R_{1.4}$ values above or below the $\pm1\sigma$ confidence interval for the EOS sample, respectively. Our results are shown in Fig.~\ref{fig:SNRfig}.

Figure~\ref{fig:SNRfig} shows that the single-parameter UR systematically underpredicts SNR for stiff EOSs and overpredicts the SNR for soft EOSs. This effect has a magnitude of $\sim 0.5-0.7$, amounting to a $\sim 5-10\%$ error in the prediction of the SNR. For comparison, a glitch magnitude that was $\sim20\%$ larger or smaller would yield the same difference in SNR. In other words, using the single-parameter, average-density UR in this case systematically under- or over-predicts the observability of these glitches, by a degree comparable to a 20\% change in the size of the glitch itself. In contrast, we find that with the addition of the two stiffness-correction terms, the 3-parameter UR recreates nearly exactly the true SNR distribution.

\section{Discussion and Conclusions} \label{sec:discussion}
In this paper, we use a large, parametric EOS sample to examine universal relations for the \textit{f}-mode frequency and damping time. On the whole, our larger sample reveals more EOS-sensitivity in these initial, single-parameter relations than the small, tabulated or polytropic EOS samples of prior works. Additionally, we observe that the residual deviation from universality is correlated with $R_{1.4}$ and $R_{1.4}/R_{1.8}$ in several relations (with $R^2\sim0.7$), including $f$-mode frequency relations with average density or effective compactness $\eta$, along with the damping time fit dependent on compactness. This same correlation is largely absent from any of the tidal deformability relations. We find that we can improve upon the standard, single-parameter URs by adding correction terms that depend on $R_{1.4}$ and $R_{1.4}/R_{1.8}$ to the UR formulae, which successfully eliminate the stiffness correlations in the UR. This is evidenced by a general reduction in the 1-$\sigma$ and 2-$\sigma$ residuals across masses, as well as in the full-sample and $1.4\Ms$ sub-sample average 1-$\sigma$ residuals.

We choose $R_{1.4}$ and the slope $R_{1.4}/R_{1.8}$ empirically, as parameters that
are related to the EOS stiffness at moderately high densities (as shown in Fig.~\ref{fig:R14_slope}). These are convenient parameters to add to the UR because they can be inexpensively estimated via standard TOV calculations and they are potentially observable quantities, thus future observational constraints on these parameters can be folded into a UR prediction.
The fact that some URs are improved by adding these correction terms suggests that the $f$-mode depends on the EOS at higher densities than can be accounted for by, e.g., the average density alone. This is consistent with past work by Ref.~\cite{Lioutas:2021jbl}, which showed that $f$-mode URs involving $\Lambda$ tended to be more accurate than those involving the radius, and hypothesized that the difference is a result of the radius' greater sensitivity to the low-density EOS. They showed that the universality could be improved with a new radius parameter, $R^{90\%}$, that encompasses the inner 90\% of the NS and thus excludes the low-density crust \cite{Lioutas:2021jbl}. A similar effect is accomplished with our corrections that depend on the radii of intermediate-to-high mass NSs.

The $f$-mode frequencies of static isolated NSs, as we explore in this work, have also been shown to be strongly correlated with the peak frequency of the post-merger GWs that are emitted from the remnant after a binary NS merger \cite{Chakravarti:2019sdc,Lioutas:2021jbl}. Numerical simulations reveal that the post-merger peak-frequency is correlated with the NS radius at fixed mass \cite[e.g.,][]{Bauswein:2012ya,Bauswein:2011tp,Takami:2014zpa,Bernuzzi:2015rla,Vretinaris:2019spn}; but that this relation breaks down as the mass-radius slope is systematically varied, also implying a dependence on the EOS at higher densities \cite{Raithel:2022orm}. Altogether, these results point toward a cohesive picture of a dependence of the $f$-modes on EOS densities that cannot be captured through low-mass radius parameters alone.

Ultimately, we find the largest residuals in URs that relate the $f$-mode frequency or damping time to combinations of the NS mass and radius, and we find significant correlations between the residuals to these URs and the two stiffness parameters. For example, for the relation between $f$-mode frequency and average-density, the average $1\sigma$ residuals are reduced from $\sim$4\% with the single-parameter UR, down to 1.5\%, with the 3-parameter relation. We find that the effective compactness and \textit{f}-Love relations have the smallest scatter among the \textit{f}-mode frequency URs; but that, even within the sub-percent residuals of the $\eta - M\omega$ relation, there are significant correlations with $R_{1.4}$ and the mass-radius slope. By introducing the two stiffness correction terms, we are able to reduce the average residuals of the 3-parameter $\eta- M\omega$ relation down to 0.06\%. In contrast, the $f$-Love relation similarly has sub-percent residuals ($\sim$0.3\%), but has minimal correlation with EOS stiffness.

To demonstrate the impact of the residual stiffness correlation in the original (i.e., uncorrected) URs, we investigate the predicted SNR of a GW signal powered by an $f$-mode that has been triggered by a Vela-like pulsar glitch. We observe that for the stiffest and softest EOSs in our sample, the peak of the SNR distribution is either underestimated or overestimated by $\sim0.5$, respectively, using the single-parameter $\sqrt{M/R^3}$ relation. This is the same magnitude effect as changing the glitch size by $\pm20\%$. 
On the other hand, our three-parameter corrected UR generates an SNR distribution that is practically indistinguishable from the distribution that uses the full-GR calculations of the $f$-modes. This suggests that using the existing average-density relation to make predictions about observability will lead to systematic over- or underestimates for particularly soft or stiff EOSs, respectively. Thus, past estimates of observability may be overly optimistic for very soft models like quark stars, and rather over-conservative for very stiff models. However, including our simple, characteristic correction terms successfully removes this bias.

Selecting which UR to use depends largely on the application. For example, some URs have been proposed as a means of testing general relativity (GR), as the UR may follow a different curve in modified theories of gravity \cite[e.g.,][]{iloveq}. Such tests require the UR to be highly insensitive to the EOS; otherwise, scatter in the relation due to the uncertainties in the unknown nuclear physics may overwhelm any modifications caused by deviations from GR. Our work points to the $\eta-M\omega$ relation as a good candidate for this sort of test, as it is the tightest relation among those that we explore, both in its original (single-parameter) form as well as with the stiffness corrections. Observational constraints on $\eta$ are anticipated from radio timing of the double pulsar system, for which the moment of inertia is expected to be measured at 10\% accuracy by 2030 \cite{Kramer:2021jcw,Hu:2024zra}, while constraints on the stiffness correction terms, $R_{1.4}$ and $R_{1.4}/R_{1.8}$, can be folded in from X-ray radius measurements or other astrophysical constraints on the EOS \cite[e.g.,][for a review]{Chatziioannou:2024jsr}.

On the other hand, expressing the $f$-mode UR directly in terms of the adiabatic tidal deformability, $\Lambda$, is a convenient choice for GW waveform modeling, where the $f$-Love UR can be used to reduce  dimensionality during parameter estimation  \cite[e.g.,][]{Pratten:2021pro,Pradhan:2022rxs,Williams:2022vct}. Recent work has shown that using an $f$-Love UR during parameter estimation for individual GW events can introduce a slight bias in the recovered tidal deformability towards softer values, but that this bias becomes negligible for populations of events \cite{Williams:2026jqv}. This is consistent with our $f$-Love results, for which
we find only a weak correlation with $R_{1.4}$ and a negligible correlation with the mass-radius slope. In other words, our results confirm that the $f$-Love relation is highly insensitive to the EOS.

As we look ahead to the exquisite sensitivity promised by next-generation GW detectors such as Cosmic Explorer \cite{Reitze:2019iox,Evans:2021gyd} and Einstein Telescope \cite{Punturo:2010zz,ET:2025xjr} and the EOS constraints this will enable \cite{Chatziioannou:2021tdi,Pacilio:2021jmq,Finstad:2022oni,Gupta:2022qgg,Iacovelli:2023nbv,Pradhan:2023zor,Walker:2024loo}, the question of UR calibration will become increasingly important, in order to avoid introducing systematic biases when marginalizing over errors in the URs \cite{Kashyap:2022wzr,Suleiman:2024ztn}. The present work represents one step forward in understanding the potential biases and limitations of the $f$-mode URs for this purpose.

We note that everything presented in this work has decomposed the perturbations to static, non-rotating NSs. Additional work might explore correlations in the URs of rotating stars, for which dynamical simulations will be necessary. The analysis in the present study could also be extended to other quasinormal modes, such as the $p-$ or $g-$ modes. We leave such extensions to future work.

\begin{acknowledgments}
We would like to thank Wynn Ho, Athul Kunjipurayil, and Tianqi Zhao for helpful discussions pertaining to this work. IC gratefully thanks Swarthmore College for financial support via the Niquette-Pohl Summer Opportunity Fund and the Sigma Xi Award. The computation for this work was performed on the Firebird HPC which is supported by Swarthmore College and Lafayette College.
\end{acknowledgments}

\bibliography{main,inspire}
\bibliographystyle{apsrev4-1}

\appendix

\section{Additional Damping Time Relations}
\label{sec:appendixdamp}

In this appendix, we explore several additional universal relations for the $f-$mode damping time. These generally perform similarly to the $M/\tau$ relations with compactness or $\Lambda^{-1/5}$ explored in Sec.~\ref{sec:damping}, but with some differences in their residual sensitivity to the EOS stiffness, which we discuss in the following.

\subsection{Existing Dimensionally-Motivated Relations} \label{sec:existingdimless}
We start by investigating URs for two different ``characteristic" damping times, motivated by the quadrupole radiation formula for GWs. In particular, the quadrupole radiation formula, together with the natural timescale for the oscillations $\omega \sim \sqrt{M/R^3}$, can be used to derive that $\tau \sim R\big( \frac{R}{M}\big)^3$ \cite{andersson,Lau:2009bu}. This leads to URs of the form

\begin{subequations}
\begin{align}
    \frac{R^4}{M^3\tau} &= a_0+a_1\frac{M}{R} \label{eq:dimless_tau_c_lin}\\
    \frac{R^4}{M^3\tau} &= a_0 + a_1\frac{M}{R} + \cdots + a_4\Big(\frac{M}{R}\Big)^4, \label{eq:dimless_tau_c}
\end{align}
\end{subequations}where eq.~(\ref{eq:dimless_tau_c_lin}) represents the initial relation proposed in \cite{andersson}, which we expand to quartic order in eq.~(\ref{eq:dimless_tau_c}). 
With their sample of 9 tabulated EOSs, Ref.~\cite{Chirenti:2015dda} reports relative errors for this UR (at linear order) that are largely below $20\%$.

\begin{table}[h]
\centering
\begin{tabular}{c|c|c!{\vrule width 1.5pt}c|c|c}
    \multicolumn{3}{c}{\textbf{Base $M/R - R^4/(M^3\tau)$}} & \multicolumn{3}{c}{\textbf{w/ Cross Terms}} \\ \noalign{\hrule height 1.3pt}
     \# Params. & $R_{1.4}$ $R^2$ & $\frac{R_{1.4}}{R_{1.8}}$ $R^2$ & \# Params. & $R_{1.4}$ $R^2$ & $\frac{R_{1.4}}{R_{1.8}}$ $R^2$ \\ \noalign{\hrule height 1.3pt}
    1 & 0.679 & 0.699 & 1 & 0.679 & 0.699  \\
    2 & 0.0754 & 0.464 & 2 w/ cr. & 0.0534 & 0.126 \\ 
    3 & 0.122 & 0.187 & 3 w/ cr. & 0.0837 & 0.00136 \\
\end{tabular}
\caption{The coefficient of determination between $R_{1.4}$ or slope and the residuals for the compactness fits to the characteristic damping time, with residuals taken at values corresponding to a $1.4 \Ms$ SLy4 NS. Note that the ``x w/ cross" fits are the second round of fits to the compactness UR that include cross terms -- the second-order terms comprising both one factor of the compactness and of a stiffness term.}
\label{tab:compactdimlessrsquared}
\end{table}

\begin{table}[h]
\centering
\begin{tabular}{c|c|c!{\vrule width 1.5pt}c|c|c}
     \multicolumn{3}{c}{\textbf{$\eta - I^2/(M^5\tau)$}} & \multicolumn{3}{c}{\textbf{$\Lambda^{-1/5} - \ln(\Lambda^{4/5}M/\tau)$}} \\ \noalign{\hrule height 1.3pt}
     \# Params. & $R_{1.4}$ $R^2$ & $\frac{R_{1.4}}{R_{1.8}}$ $R^2$ & \# Params. & $R_{1.4}$ $R^2$ & $\frac{R_{1.4}}{R_{1.8}}$ $R^2$ \\ \noalign{\hrule height 1.3pt}
    1 & 0.392 & 0.221 & 1 & 0.116 & 0.104 \\
    2 & 0.0223 & 0.0281 & 2 & 0.00745 & 0.00195 \\
    3 & 0.0230 & 0.0136 & 3 & 0.00682 & 0.0152 \\
\end{tabular}
\caption{The coefficient of determination between $R_{1.4}$ or slope and the residuals for the $\eta$ and $\Lambda^{-1/5}$ fits to the characteristic damping time, with residuals taken at values corresponding to a $1.4 \Ms$ SLy4 NS. }
\label{tab:dimlesstaursquared}
\end{table}

We also investigate an effective compactness UR, as introduced in Ref.~\cite{Lau:2009bu}, given by
\begin{subequations}
\begin{align}
    \frac{I^2}{M^5\tau} &= a_0 + a_2\eta^2 \label{eq:dimless_tau_eta_lin} \\ 
    \frac{I^2}{M^5\tau} &= a_0 + a_1\eta + \cdots + a_4\eta^4, \label{eq:dimless_tau_eta}
\end{align}
\end{subequations}where $I^2/(M^5\tau)$ is another characteristic damping time. As for the previous relation, eq.~(\ref{eq:dimless_tau_eta_lin}) is the initial proposed fit, which we expand in eq.~(\ref{eq:dimless_tau_eta}). The initial sample of $\sim$10 tabulated EOSs in Ref.~\cite{Lau:2009bu} yields fractional errors in this UR between $1.1-2.2\%$, while the slightly larger sample in Ref.~\cite{Chirenti:2015dda} finds relative errors of $\sim3-5\%$.
\begin{figure}
    \centering
    \includegraphics[width=0.95\linewidth]{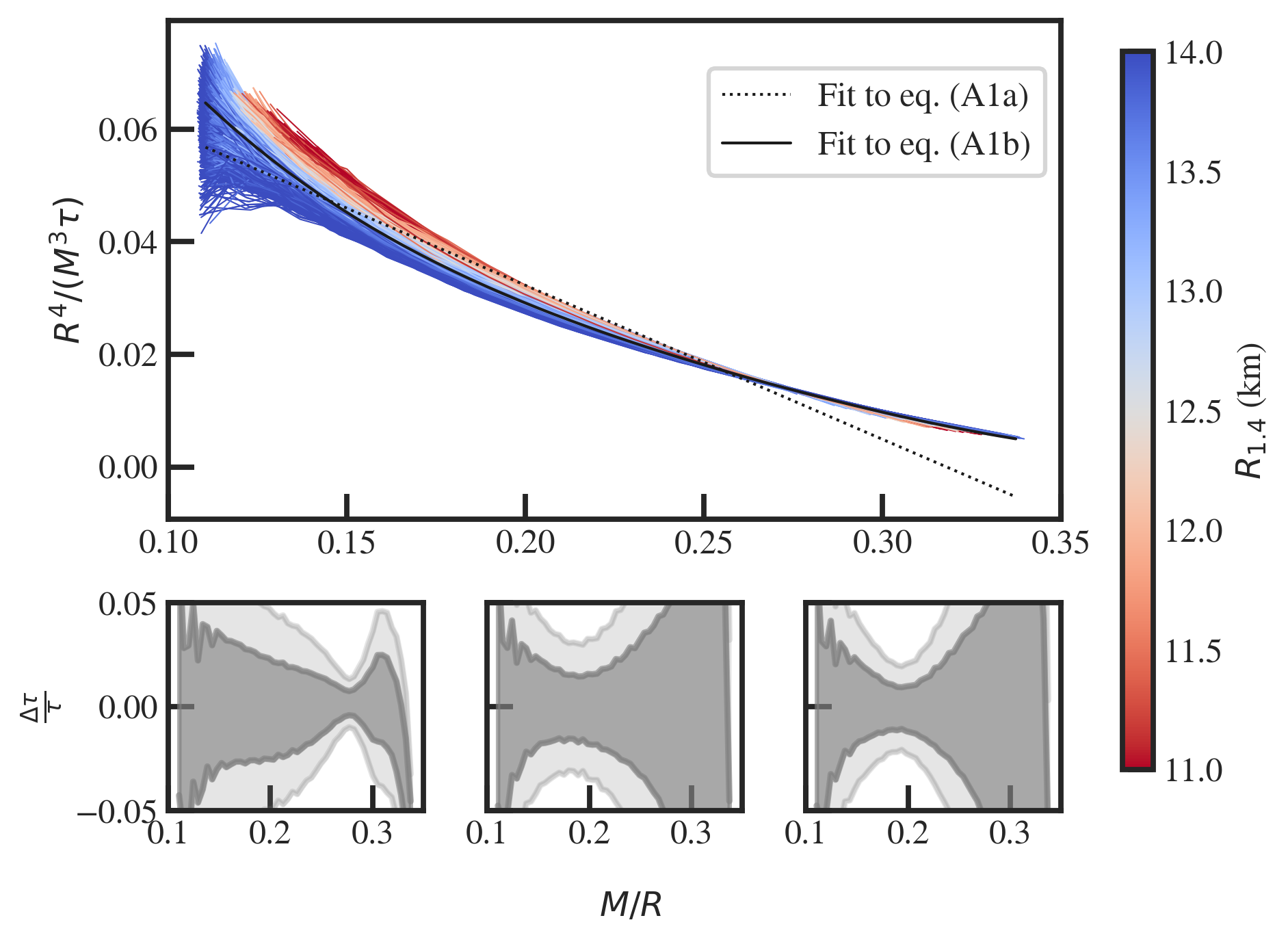}
    \caption{Compactness UR for the dimensionless damping time $R^4/(M^3\tau)$. The 3 subplots below display the 1-$\sigma$ and $2\sigma$ values for the residuals of quartic fits in $M/R$. From left to right: the base (single-parameter) relation, with a linear correction term in $R_{1.4}$, and with linear correction terms in both $R_{1.4}$ and the slope. Observe that the correction terms appear to generally broaden the residuals in the low- or high-$M/R$ areas.}
    \label{fig:compactdamping}
\end{figure}

\begin{figure}
    \centering
    \includegraphics[width=0.95\linewidth]{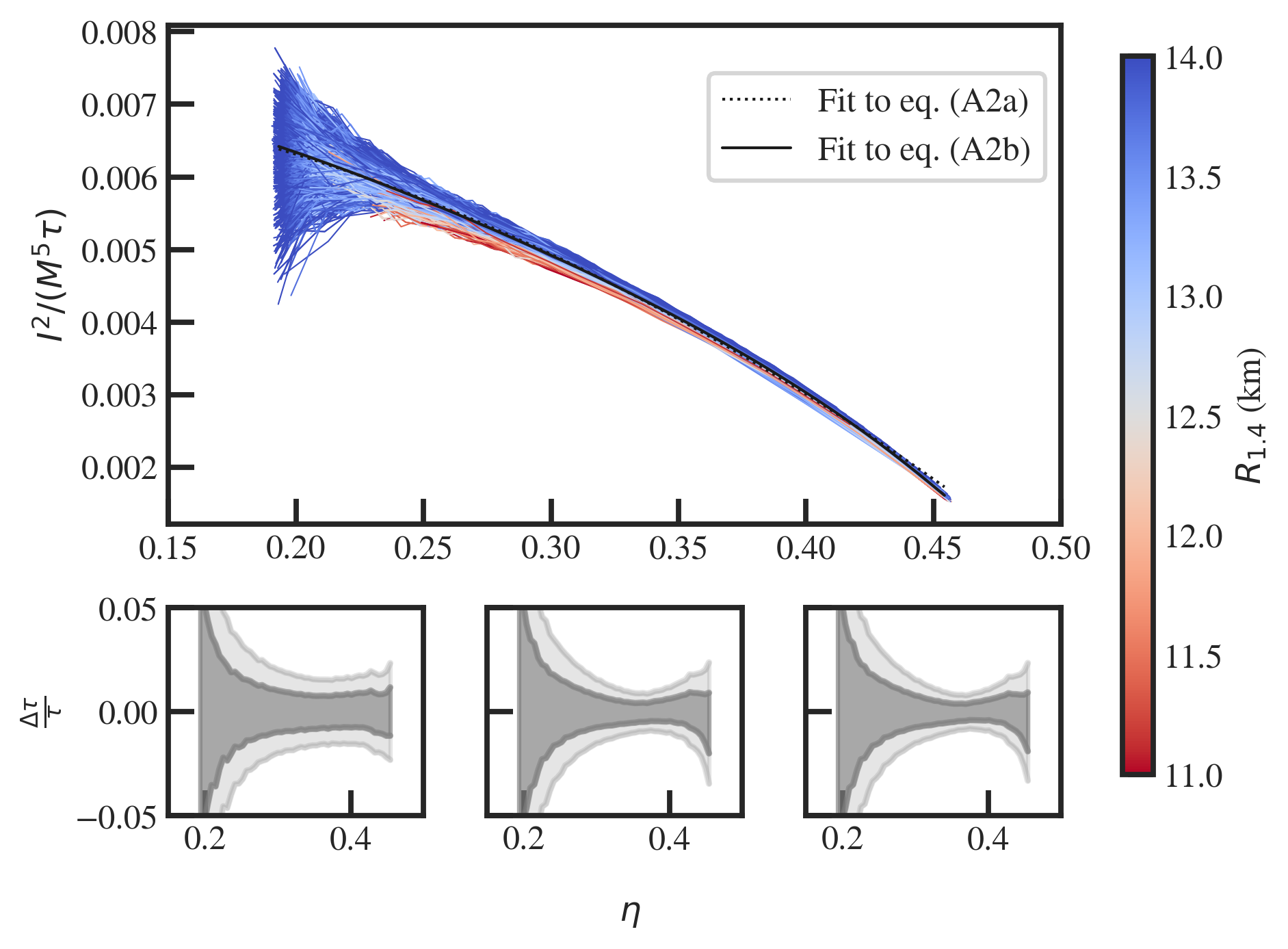}
    \caption{$\eta$ UR for the dimensionless damping time $I^2/(M^5\tau)$. The residual plots are as in Figure \ref{fig:compactdamping}. Evidently, the residuals for each of the fits to $I^2/(M^5\tau)$ are better than any of the compactness fits. Independently, the correction terms do improve the residuals as more are added, although the effect largely manifests as a tightening of the relation for intermediate $\eta$ values. }
    \label{fig:etadamping}
\end{figure}

We fit both of these characteristic damping time relations (eqs.~\ref{eq:dimless_tau_c} and \ref{eq:dimless_tau_eta}) with our large EOS sample, and show the results in Figs. \ref{fig:compactdamping} and \ref{fig:etadamping}, with best-fit coefficients in Tables \ref{tab:compactnesscombo} and \ref{tab:etacombo}, and residuals in Table \ref{tab:dimtauresid}. In the compactness relation, 
we find residuals that are consistently below $\sim5\%$, whereas the $\eta$ version of the relation yields $\sim1\%$ residuals (although they get as large as $7\%$ in the low-$\eta$/low-mass regime). 

For the single-parameter fit to compactness (eq. \ref{eq:dimless_tau_c}, shown in Fig.~\ref{fig:compactdamping}), we find significant residual correlations with both $R_{1.4}$ and the mass-radius slope ($R^2\sim0.7$).\footnote{We note that the exact correlation coefficients are unchanged from the values found for the $M/\tau$ URs in Sec.~\ref{sec:damping}, because calculating the $\Delta\tau/\tau$ residual entails removing the same factors that distinguish $M/\tau$ from the characteristic damping time in the first place.} However, when we apply the linear $R_{1.4}$ and slope correction terms to the compactness relation, we see residuals actually increase in size at small or large compactness values (evidenced by a $27-30\%$ increase in the full-sample 1-$\sigma$ value; see Table~\ref{tab:dimtauresid} and the bottom right panels of Fig.~\ref{fig:compactdamping}). Despite this, the fit is constrained better for the $1.4 \Ms$ NSs,  with the 3-parameter fit experiencing a $43\%$ reduction in the $1.4\Ms$ 1-$\sigma$ residual.

Unlike for the other URs we have explored, Figure~\ref{fig:compactdamping} shows that the trend in the residual EOS-sensitivity with $R_{1.4}$ \textit{reverses direction} between high and low compactness values, for this combination of parameters. This explains why linearly removing a dependence on $R_{1.4}$ causes the residuals to increase at certain values of $M/R$. To account for this, we introduce two second-order correction terms, so that the overall fit is now given by
\begin{multline}
    \frac{R^4}{M^3\tau} = a_0 + a_1\Bigg(\frac{M}{R}\Bigg) + \cdots + a_4\Big(\frac{M}{R}\Big)^4 +\\
     b_1 R_{1.4} + c_1\Bigg(\frac{R_{1.4}}{R_{1.8}}\Bigg) + \\ 
     d_2 R_{1.4}\Bigg(\frac{M}{R}\Bigg) +
    e_2\Bigg(\frac{R_{1.4}}{R_{1.8}}\Bigg)\Bigg(\frac{M}{R}\Bigg)
\end{multline}

\begin{figure}
    \centering
    \includegraphics[width=0.95\linewidth]{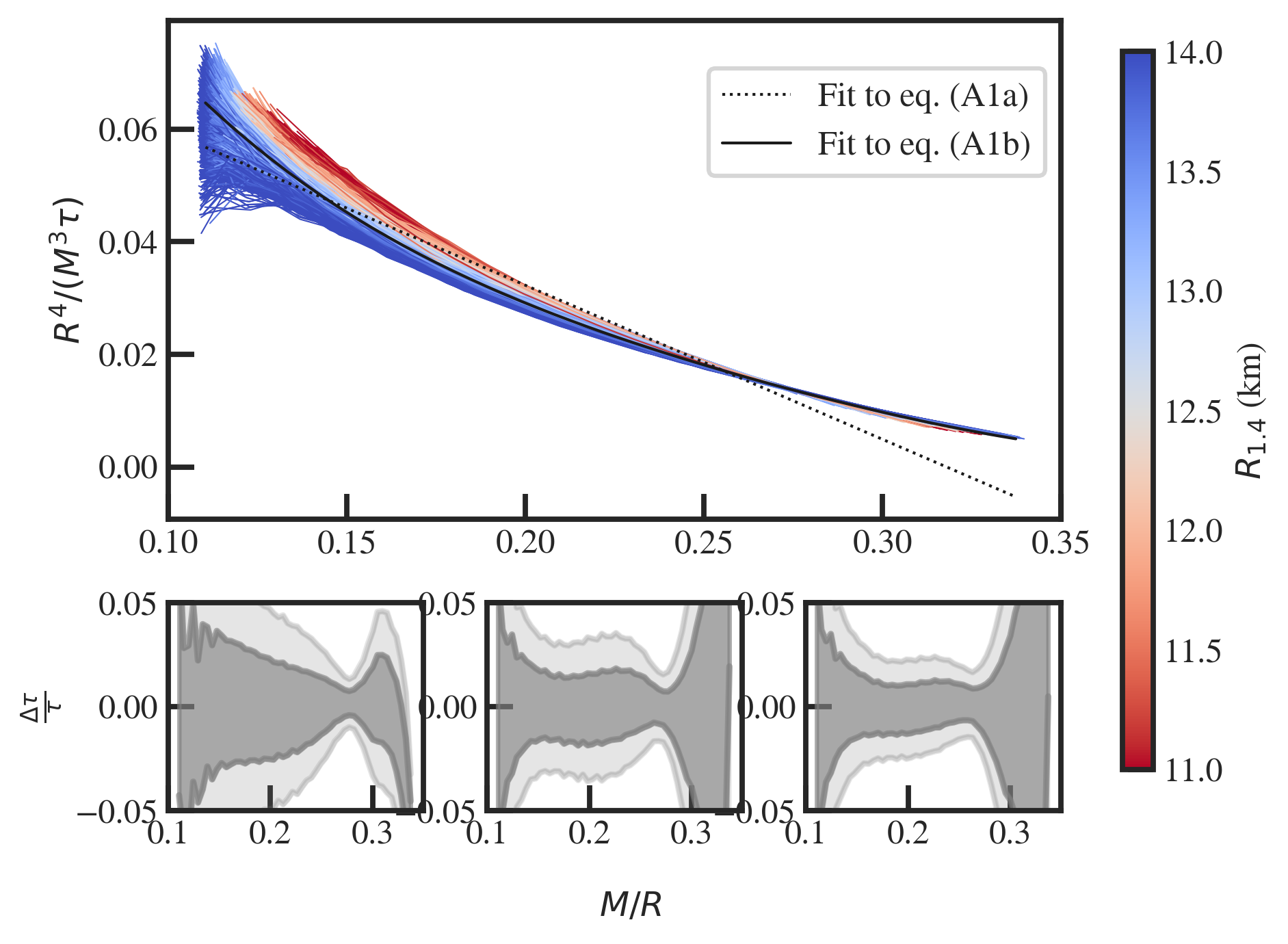}
    \caption{Same as Fig.~\ref{fig:compactdamping}, but with the cross correction-term proportional to $(R_{1.4})(\frac{M}{R})$ added to the second residual plot and the cross-correction term proportional to $(\frac{R_{1.4}}{R_{1.8}})(\frac{M}{R})$ added to the third residual plot. These terms are meant to address the fact that the $R_{1.4}$ correlation remains throughout the sample, but the trend is opposite at low and high ends of $M/R$ space. This fit behaves better than those in Fig.~(\ref{fig:compactdamping}), but still suffers from large residuals at high $M/R$ values. }
    \label{fig:mixcorrectcompact}
\end{figure}

Figure~\ref{fig:mixcorrectcompact} shows that the residuals to the 3-parameter fit with these cross-terms are smaller than those of the standard 3-parameter fit in Fig.~\ref{fig:compactdamping}, reducing residuals within $0.15 < M/R < 0.3$ to $\sim$2\%. However, large residuals remain at the highest values of $M/R$. The full-sample and $1.4\Ms$ residuals are reduced by $\sim10\%$ and $\sim 50\%$ of their single-parameter values for the 3-parameter fits with cross terms (see Table \ref{tab:dimtauresid}). We report the complete fit parameters in Tables \ref{tab:compactnesscombo} and \ref{tab:compactextra}.

We find that the residuals to the single-parameter UR with $\eta$ (eq. \ref{eq:dimless_tau_eta}, shown in Fig.~\ref{fig:etadamping}), are less correlated with $R_{1.4}$ and the slope, with $R^2$ values of $0.39$ and $0.22$, respectively. As a result, we do not find significant differences in the residuals with the addition of the stiffness corrections terms. There is a small tightening of the relation around $0.3<\eta<0.4$ region with the addition of the correction terms (see the bottom row of Fig.~\ref{fig:etadamping}), which is mirrored by a reduction of the $1.4\Ms$ sample residual by $\sim10\%$ of its uncorrected value. However, the full-sample 1-$\sigma$ mark is reduced by only $\sim5\%$ of its single-parameter value with the stiffness correction terms.

Comparing to the fits for the mass-scaled damping time $(M/\tau)$ in Sec.~\ref{sec:damping}, we thus see the additional factors of $M,R,$ or $I$ that enter into the characteristic damping time do not affect the overall residuals for the single-parameter fits (see Tables \ref{tab:tauresid} and \ref{tab:dimtauresid}). However, they do result in less EOS-sensitivity (i.e., smaller residuals) for most of the multi-parameter fits.

\subsection{Analogous $\Lambda$ Relation}
\begin{figure}
    \centering
    \includegraphics[width=0.95\linewidth]{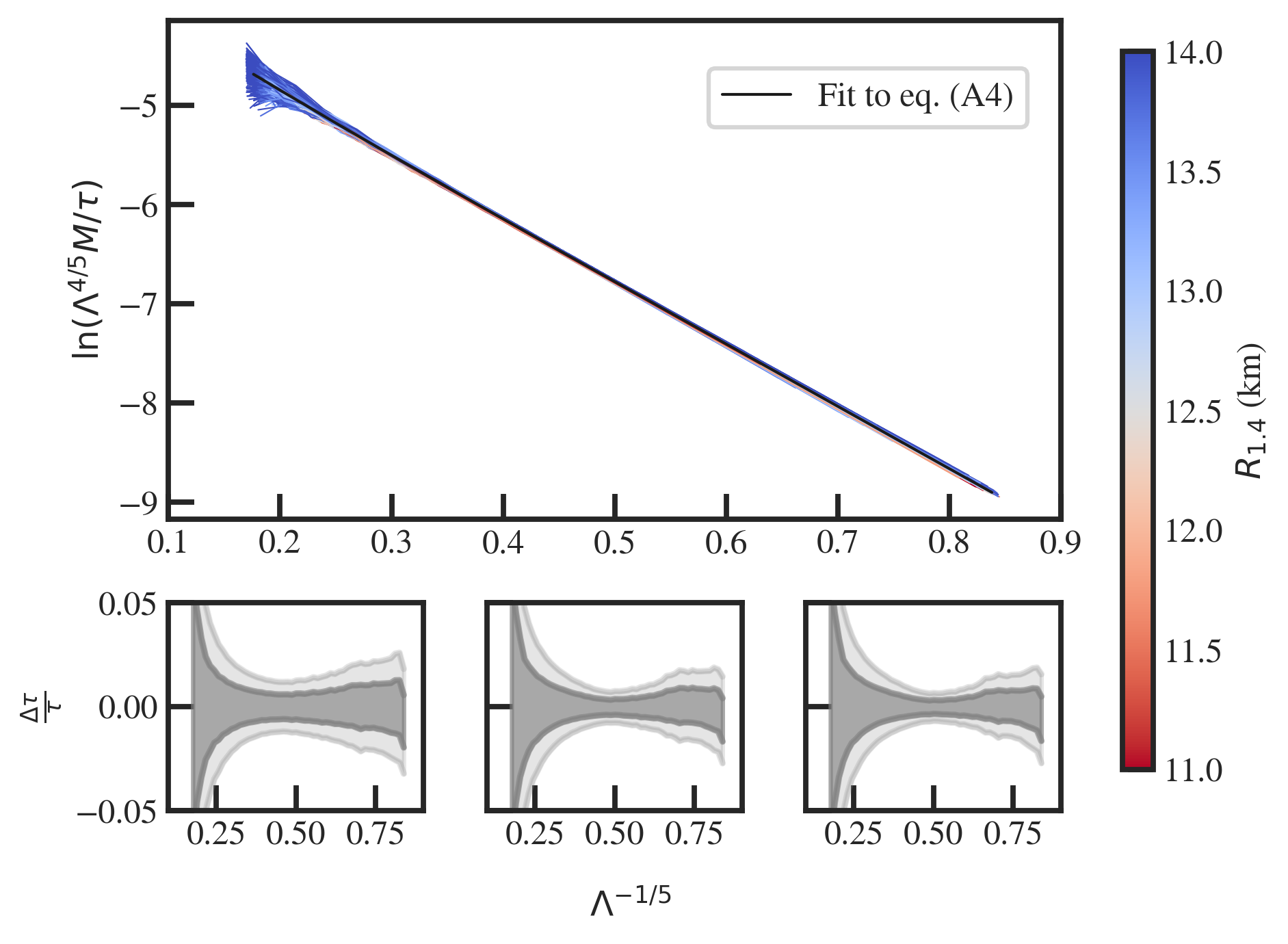}
    \caption{Reduced tidal deformability UR for the dimensionless damping time $\ln(\frac{\Lambda^{4/5} M}{\tau})$. The residual plots are as in Fig.~\ref{fig:compactdamping}. As expected, the residuals are very similar to the $\eta$ fits that are presented in Fig.~\ref{fig:etadamping}. Most of the spread in the relation occurs at the low end of the mass spectrum, but the UR constrains the rest of the mass spectrum to $<1\%$ residuals. }
    \label{fig:lambdamping}
\end{figure}
We also introduce a $\Lambda$-dependent UR for a new characteristic damping time. Keeping the $f$-mode parameters (both frequency and damping time) directly in terms of the tidal deformability is useful for multiple reasons, both for waveform fitting and because GW observations are predicted to be far more numerous than moment of inertia measurements. Our new characteristic combination is again motivated by the fact that $\eta \sim \Lambda^{-1/5}$, via the I-Love-Q compactness relations \cite{iloveq}. Applying this transformation to the $\eta - I^2/(M^5\tau)$ relation and taking a logarithm to admit a polynomial fit, we arrive at a tidal deformability fit for the characteristic damping time:
\begin{equation}\label{dimlesslambtau}
    \ln(\Lambda^{4/5}M/\tau) = a_0+a_1x+a_2x^2+a_3x^3+a_4x^4,
\end{equation}
where $x=\Lambda^{-1/5}$.  

\begin{table}[h]
\centering
\begin{tabular}{!{\vrule width 1.5pt}c|c|c|c|c|c!{\vrule width 1.5pt}}
    \noalign{\hrule height 1.5pt}
     & \multirow{2}{*}{\# Params.} & Avg. 1-$\sigma$ & \multirow{2}{*}{\% Red.} & $1.4\Ms$ 1-$\sigma$ & \multirow{2}{*}{\% Red.} \\
     & & ($10^{-2}$) & & ($10^{-2}$) & \\ \noalign{\hrule height 1.5pt}
    \multirow{5}{1.5cm}{$\frac{M}R-\frac{R^4}{(M^3\tau)}$} & 1 & 2.83 & --- & 2.96 & --- \\ \cline{2-6} 
        & 2 & 3.67 & -29.7 & 1.88 & 36.5 \\ \cline{2-6} 
       & 3 & 3.60 & -27.2 & 1.69 & 42.9 \\
       \cline{2-6} & 2 w/ cr. & 2.56 & 9.5 & 1.61 & 45.6 \\
       \cline{2-6} & 3 w/ cr. & 2.53 & 10.6 & 1.40 & 52.8 \\ \noalign{\hrule height 1.5pt}
    \multirow{3}{1.5cm}{$\eta-\frac{I^2}{(M^5\tau)}$} & 1 & 1.82 & --- & 1.38 & --- \\ \cline{2-6} 
     & 2 & 1.73 & 4.9 & 1.24 & 10.2 \\ \cline{2-6} 
     & 3 & 1.72 & 5.2 & 1.24 & 10.2 \\ \noalign{\hrule height 1.5pt}
    \multirow{3}{1.5cm}{$\Lambda^{-1/5} -\ln(\frac{\Lambda^{4/5}M}{\tau})$} & 1 & 1.81 & --- & 1.61 & --- \\ \cline{2-6} 
     & 2 & 1.77 & 2.3 & 1.56 & 3.3 \\ \cline{2-6} 
     & 3 & 1.76 & 2.9 & 1.53 & 4.9 \\ \noalign{\hrule height 1.5pt}
\end{tabular}
\caption{Residuals for \textit{f}-mode dimensionless damping time URs. We report the average 1-$\sigma$ residuals for both the full sample and the sub-sample of $1.4\Ms$ NSs, along with the percent reductions in magnitude of the residuals for the multi-parameter fits, compared to the single-parameter fits. The ``x w/ cr." fits are the second round of fits to the compactness UR that include cross terms -- i.e., the second-order terms that include one factor of the relevant stiffness correction term and one factor of the compactness. Additionally, the negative values for the 2- and 3-parameter fits to the compactness relation indicate an increase in the size of residual.}
\label{tab:dimtauresid}
\end{table}

This characteristic damping time fit has minimal correlations with $R_{1.4}$ and the slope ($R^2=0.12$ and 0.10, respectively), matching the $M/\tau$ fit results -- as expected.

Also, due to the standard I-Love universality between $\eta$ and $\Lambda$, this fit performs nearly identically to the $\eta$ fit for the characteristic damping time in eq.~(\ref{eq:dimless_tau_eta}). We find $1.4\Ms$ 1$\sigma$ residuals of $\sim1.5\%$ for both the $\eta$ and $\Lambda^{-1/5}$ relations, with or without the stiffness correction terms (see Table \ref{tab:lambdacombo} for fit parameters and \ref{tab:dimtauresid} for residuals). In both cases, the addition of the EOS-dependent correction terms make only very marginal ($\leq\mathcal{O}(0.1\%)$) improvements. In particular, the tidal deformability fit suffers from the same EOS-sensitivity at low masses as the effective compactness fit (see discussion in Sec. \ref{sec:existingdimless}) and likewise exhibits general consistency throughout the rest of the mass spectrum. Correction terms for this UR between $\Lambda^{-1/5}$ and the characteristic damping time are more successful than those for the corresponding $M/\tau$ fit, succeeding in improving the full-sample residuals by $\sim3\%$ and the $1.4\Ms$ sub-sample measure by $\sim5\%$; although we note that in both cases, the residual dependence on EOS stiffness in the $\Lambda^{-1/5}$ URs is quite small.

\section{Fit Parameters to URs}
\label{sec:appendixparams}
Here, we report the best-fit parameters for all URs that we explore in this work: the average density, $\eta$, and tidal deformability relations for frequency, mass-scaled damping time, and characteristic damping time.

\begin{table*}[t]
    \centering
    \renewcommand{\arraystretch}{1.5}
    \footnotesize
    \begin{tabular}{c|c|c|c|c|c|c|c}
          & $a_0$ & $a_1$ & $a_2$ & $a_3$ & $a_4$ & $b_1$ & $c_1$ \\ \hline
         $\sqrt{\frac{M}{R^3}}$ base fit & $-1.4546\times10^0$ & $2.7168\times10^2$ & $-1.0055\times10^4$ & $1.7860\times10^5$ & $-1.1192\times10^6$ & --------- & --------- \\ \hline
         $R_{1.4}$ only & $2.6962\times10^0$ & $-9.8713\times10^0$ & $7.6377\times10^2$ & $-3.0245\times10^3$ & $-2.0247\times10^4$ & $-1.0663\times10^{-1}$ & ---------  \\ \hline
         $R_{1.4}$ and $\frac{R_{1.4}}{R_{1.8}}$ & $9.978\times10^{-1}$ & $-3.2748\times10^1$ & $1.6799\times10^3$ & $-1.811\times10^4$ & $6.3674\times10^4$ & $-7.7815\times10^{-2}$ & $1.5415\times10^0$
    \end{tabular}
    \caption{Best-fit coefficients for the average density fits to $f$-mode frequency: the single-parameter quartic fit in average density (eq.~\ref{uncorrectedquartic}), the quartic fit in average density plus a linear $R_{1.4}$ correction term (eq.~\ref{r1point4quartic}), and the quartic fit in average density plus linear correction terms in $R_{1.4}$ and $R_{1.4}/R_{1.8}$ (eq.~\ref{correctedquartic}). The coefficients $a_n$ have units of kHz $\text{km}^n$, whereas $b_1$ has units of kHz $\text{km}^{-1}$ and $c_1$ has units of kHz.  } 
    \label{tab:coeffs}
\end{table*}

\begin{table*}[t]
    \centering
    \renewcommand{\arraystretch}{1.5}
    \scriptsize
    \begin{tabular}{c|c|c|c|c|c|c|c|c}
         & $a_0$ & $a_1$ & $a_2$ & $a_3$ & $a_4$ & $a_5$ & $b_1$ & $c_1$ \\ \hline
         $\eta$ base fit & $3.7917\times10^{-2}$ & $-6.1077\times10^{-1}$ & $5.7452\times10^{0}$ & $-1.8032\times10^{1}$ & $3.1349\times10^{1}$ & $-2.1669\times10^{1}$ & --------- & --------- \\ \hline
         $R_{1.4}$ only & $3.2913\times10^{-2}$ & $-5.4199\times10^{-1}$ & $5.3167\times10^{0}$ & $-1.6716\times10^{1}$ & $2.9360\times10^{1}$ & $-2.0481\times10^{1}$ & $4.9413\times10^{-5}$ & --------- \\ \hline
         $R_{1.4}$ and $\frac{R_{1.4}}{R_{1.8}}$ & $3.7204\times10^{-2}$ & $-5.3768\times10^{-1}$ & $5.2921\times10^{0}$ & $-1.6650\times10^{1}$ & $2.9273\times10^{1}$ & $-2.0439\times10^{1}$ & $-2.0976\times10^{-5}$ & $-3.7032\times10^{-3}$
    \end{tabular}
    \caption{Best-fit coefficients for the $\eta$ fits to $M\omega$: the ``base" quintic fit in $\eta$, the quintic fit in $\eta$ plus a linear $R_{1.4}$ correction term, and the quintic fit in $\eta$ plus linear correction terms in $R_{1.4}$ and $R_{1.4}/R_{1.8}$. The coefficients $a_n$ and $c_1$ are all dimensionless, whereas $b_1$ has units $\text{km}^{-1}$.} 
    \label{tab:feta}
\end{table*}

\begin{table*}[t]
    \centering
    \renewcommand{\arraystretch}{1.5}
    \footnotesize
    \begin{tabular}{c|c|c|c|c|c|c|c}
         & $a_0$ & $a_1$ & $a_2$ & $a_3$ & $a_4$ & $b_1$ & $c_1$  \\ \hline
         $\ln \Lambda$ base fit & $1.8214\times10^{-1}$ & $-7.2488\times10^{-3}$ & $-4.1613\times10^{-3}$ & $5.3030\times10^{-4}$ & $-1.9478\times10^{-5}$ & --------- & --------- \\ \hline
         $R_{1.4}$ only & $1.8449\times10^{-1}$ & $-7.3571\times10^{-3}$ & $-4.1155\times10^{-3}$ & $5.2274\times10^{-4}$ & $-1.9048\times10^{-5}$ & $-1.7111\times10^{-4}$ & --------- \\ \hline
         $R_{1.4}$ and $\frac{R_{1.4}}{R_{1.8}}$ & $1.8726\times10^{-1}$ & $-7.3470\times10^{-3}$ & $-4.1189\times10^{-3}$ & $5.2325\times10^{-4}$ & $-1.9074\times10^{-5}$ & $-2.1388\times10^{-4}$ & $-2.2523\times10^{-3}$
    \end{tabular}
    \caption{Best-fit coefficients for the $\ln\Lambda$ fits to $M\omega$: the ``base" quartic fit in $\ln\Lambda$, the quartic fit in $\ln\Lambda$ plus a linear $R_{1.4}$ correction term, and the quartic fit in $\ln\Lambda$ plus linear correction terms in $R_{1.4}$ and $R_{1.4}/R_{1.8}$. The coefficients $a_n$ and $c_1$ are dimensionless, whereas $b_1$ has units of $\text{km}^{-1}$.} 
    \label{tab:flove}
\end{table*}

\begin{table*}[]
    \centering
    \renewcommand{\arraystretch}{1.5}
    \footnotesize
    \begin{tabular}{c|c|c|c|c|c|c|c|c}
         & $a_0$ & $a_1$ & $a_2$ & $a_3$ & $a_4$ & $a_5$ & $b_1$ & $c_1$ \\ \hline
         $M/R$ base fit & $-2.9030 \times 10^{6}$& $8.0299 \times 10^{7}$& $-8.6670 \times 10^{8}$& $4.7874 \times 10^{9}$& $-1.1551 \times 10^{10}$& $9.6615 \times 10^{9}$ & --------- & --------- \\ \hline
         $R_{1.4}$ only & $-2.1181 \times 10^{6}$& $6.8562 \times 10^{7}$& $-7.5644 \times 10^{8}$& $4.2835 \times 10^{9}$& $-1.0429 \times 10^{10}$& $8.6850 \times 10^{9}$& $-2.2566 \times 10^{4}$ & --------- \\ \hline
         $R_{1.4}$ and $\frac{R_{1.4}}{R_{1.8}}$ & $-3.1110 \times 10^{6}$& $6.9457 \times 10^{7}$& $-7.6646 \times 10^{8}$& $4.3379 \times 10^{9}$& $-1.0571 \times 10^{10}$& $8.8289 \times 10^{9}$& $-7.7964 \times 10^{3}$& $7.7736 \times 10^{5}$ 
    \end{tabular}
    \caption{Best-fit coefficients for the $M/R$ fits to $M/\tau$: the ``base" quintic fit in $M/R$, the quintic fit in $M/R$ plus a linear $R_{1.4}$ correction term, and the quintic fit in $M/R$ plus linear correction terms in $R_{1.4}$ and $R_{1.4}/R_{1.8}$. The coefficients $a_n$ and $c_1$ all have units of $s$, whereas $b_1$ has units of $s$ $\text{km}^{-1}$.} 
    \label{tab:compactnessdamp}
\end{table*}

\begin{table*}[]
    \centering
    \renewcommand{\arraystretch}{1.5}
    \footnotesize
    \begin{tabular}{c|c|c|c|c|c|c|c}
         & $a_0$ & $a_1$ & $a_2$ & $a_3$ & $a_4$ & $b_1$ & $c_1$ \\ \hline
         $M/R$ base fit & $1.8374\times10^{-1}$ & $-1.7316\times10^{0}$ & $7.6350\times10^{0}$ & $-1.7334\times10^{1}$ & $1.5597\times10^{1}$ & --------- & --------- \\ \hline
         $R_{1.4}$ only & $2.0599\times10^{-1}$ & $-1.8785\times10^{0}$ & $8.6735\times10^{0}$ & $-2.0502\times10^{1}$ & $1.9123\times10^{1}$ & $-1.1088\times10^{-3}$ & --------- \\ \hline
         $R_{1.4}$ and $\frac{R_{1.4}}{R_{1.8}}$ & $1.8123\times10^{-1}$ & $-1.8888\times10^{0}$ & $8.7553\times10^{0}$ & $-2.0775\times10^{1}$ & $1.9459\times10^{1}$ & $-7.2144\times10^{-4}$ & $2.0395\times10^{-2}$
    \end{tabular}
    \caption{Best-fit coefficients for the $M/R$ fits to $R^4/(M^3\tau)$: the ``base" quartic fit in $M/R$, the quartic fit in $M/R$ plus a linear $R_{1.4}$ correction term, and the quartic fit in $M/R$ plus linear correction terms in $R_{1.4}$ and $R_{1.4}/R_{1.8}$. The coefficients $a_n$ and $c_1$ are all dimensionless, whereas $b_1$ has units of $\text{km}^{-1}$.} 
    \label{tab:compactnesscombo}
\end{table*}

\begin{table*}[]
    \centering
    \renewcommand{\arraystretch}{1.5}
    \footnotesize
    \begin{tabular}{c|c|c|c|c|c|c|c|c}
         & $a_0$ & $a_1$ & $a_2$ & $a_3$ & $a_4$ & $a_5$ & $b_1$ & $c_1$ \\ \hline
         $\eta$ base fit & $4.2984 \times 10^{6}$& $-8.1696 \times 10^{7}$& $6.1889 \times 10^{8}$& $-2.3486 \times 10^{9}$& $4.7703 \times 10^{9}$& $-3.9189 \times 10^{9}$ & --------- & --------- \\ \hline
         $R_{1.4}$ only & $2.7601 \times 10^{6}$& $-6.0550 \times 10^{7}$& $4.8713 \times 10^{8}$& $-1.9442 \times 10^{9}$& $4.1584 \times 10^{9}$& $-3.5535 \times 10^{9}$& $1.5195 \times 10^{4}$ & --------- \\ \hline
         $R_{1.4}$ and $\frac{R_{1.4}}{R_{1.8}}$ & $2.9803 \times 10^{6}$& $-6.0375 \times 10^{7}$& $4.8618 \times 10^{8}$& $-1.9418 \times 10^{9}$& $4.1556 \times 10^{9}$& $-3.5524 \times 10^{9}$& $1.1621 \times 10^{4}$& $-1.8771 \times 10^{5}$
    \end{tabular}
    \caption{Best-fit coefficients for the $\eta$ fits to $M/\tau$: the ``base" quintic fit in $\eta$, the quintic fit in $\eta$ plus a linear $R_{1.4}$ correction term, and the quintic fit in $\eta$ plus linear correction terms in $R_{1.4}$ and $R_{1.4}/R_{1.8}$. The coefficients $a_n$ and $c_1$ all have units of $s$, whereas $b_1$ has units of $s$ $\text{km}^{-1}$.} 
    \label{tab:etadamp}
\end{table*}

\begin{table*}[]
    \centering
    \renewcommand{\arraystretch}{1.5}
    \footnotesize
    \begin{tabular}{c|c|c|c|c|c|c|c}
         & $a_0$ & $a_1$ & $a_2$ & $a_3$ & $a_4$ & $b_1$ & $c_1$ \\ \hline
         $\eta$ base fit & $6.7248\times10^{-3}$ & $1.5875\times10^{-2}$ & $-1.3874\times10^{-1}$ & $3.0721\times10^{-1}$ & $-2.9335\times10^{-1}$ & --------- & --------- \\ \hline
         $R_{1.4}$ only & $5.2114\times10^{-3}$ & $2.7248\times10^{-2}$ & $-1.9100\times10^{-1}$ & $4.1195\times10^{-1}$ & $-3.7068\times10^{-1}$ & $4.5399\times10^{-5}$ & --------- \\ \hline
         $R_{1.4}$ and $\frac{R_{1.4}}{R_{1.8}}$ & $5.4937\times10^{-3}$ & $2.7402\times10^{-2}$ & $-1.9175\times10^{-1}$ & $4.1353\times10^{-1}$ & $-3.7190\times10^{-1}$ & $4.0881\times10^{-5}$ & $-2.3726\times10^{-4}$
    \end{tabular}
    \caption{Best-fit coefficients for the $\eta$ fits to $I^2/(M^5\tau)$: the ``base" quartic fit in $\eta$, the quartic fit in $\eta$ plus a linear $R_{1.4}$ correction term, and the quartic fit in $\eta$ plus linear correction terms in $R_{1.4}$ and $R_{1.4}/R_{1.8}$. The coefficients $a_n$ and $c_1$ are all dimensionless, whereas $b_1$ has units of $\text{km}^{-1}$.} 
    \label{tab:etacombo}
\end{table*}

\begin{table*}[]
    \centering
    \renewcommand{\arraystretch}{1.5}
    \footnotesize
    \begin{tabular}{c|c|c|c|c|c|c|c|c}
         & $a_0$ & $a_1$ & $a_2$ & $a_3$ & $a_4$ & $a_5$ & $b_1$ & $c_1$ \\ \hline
         $\Lambda^{-1/5}$ base fit & $7.6489 \times 10^{5}$& $-1.2479 \times 10^{7}$& $7.3749 \times 10^{7}$& $-1.1911 \times 10^{8}$& $7.0969 \times 10^{7}$& $-1.2018 \times 10^{7}$ & --------- & --------- \\ \hline
         $R_{1.4}$ only & $5.5302 \times 10^{5}$& $-1.1643 \times 10^{7}$& $6.9961 \times 10^{7}$& $-1.1093 \times 10^{8}$& $6.2514 \times 10^{7}$& $-8.6565 \times 10^{6}$& $1.0673 \times 10^{4}$ & --------- \\ \hline
         $R_{1.4}$ and $\frac{R_{1.4}}{R_{1.8}}$ & $7.9324 \times 10^{5}$& $-1.1668 \times 10^{7}$& $7.0103 \times 10^{7}$& $-1.1132 \times 10^{8}$& $6.3014 \times 10^{7}$& $-8.8941 \times 10^{6}$& $7.0026 \times 10^{3}$& $-1.9274 \times 10^{5}$
    \end{tabular}
    \caption{Best-fit coefficients for the $\Lambda^{-1/5}$ fits to $M/\tau$: the ``base" quintic fit in $\Lambda^{-1/5}$, the quartic fit in $\Lambda^{-1/5}$ plus a linear $R_{1.4}$ correction term, and the quartic fit in $\Lambda^{-1/5}$ plus linear correction terms in $R_{1.4}$ and $R_{1.4}/R_{1.8}$. The coefficients $a_n$ and $c_1$ all have units of $s$, whereas $b_1$ has units of $s$ $\text{km}^{-1}$.} 
    \label{tab:lambdadamp}
\end{table*}

\begin{table*}[]
    \centering
    \renewcommand{\arraystretch}{1.5}
    \footnotesize
    \begin{tabular}{c|c|c|c|c|c|c|c}
         & $a_0$ & $a_1$ & $a_2$ & $a_3$ & $a_4$ & $b_1$ & $c_1$ \\ \hline
         $\Lambda^{-1/5}$ base fit & $-3.4398\times10^{0}$ & $-7.4950\times10^{0}$ & $2.7850\times10^{0}$ & $-2.8272\times10^{0}$ & $1.0729\times10^{0}$ & --------- & --------- \\ \hline
         $R_{1.4}$ only & $-3.5399\times10^{0}$ & $-7.3641\times10^{0}$ & $2.3576\times10^{0}$ & $-2.2393\times10^{0}$ & $7.8445\times10^{-1}$ & $6.4748\times10^{-3}$ & --------- \\ \hline
         $R_{1.4}$ and $\frac{R_{1.4}}{R_{1.8}}$ & $-3.3486\times10^{0}$ & $-7.3488\times10^{0}$ & $2.2963\times10^{0}$ & $-2.1438\times10^{0}$ & $7.3260\times10^{-1}$ & $3.5129\times10^{-3}$ & $-1.5564\times10^{-1}$ 
    \end{tabular}
    \caption{Best-fit coefficients for the $\Lambda^{-1/5}$ fits to $\ln(\Lambda^{4/5}M/\tau)$: the ``base" quartic fit in $\Lambda^{-1/5}$, the quartic fit in $\Lambda^{-1/5}$ plus a linear $R_{1.4}$ correction term, and the quartic fit in $\Lambda^{-1/5}$ plus linear correction terms in $R_{1.4}$ and $R_{1.4}/R_{1.8}$. The coefficients $a_n$ and $c_1$ are all dimensionless, whereas $b_1$ has units of $\text{km}^{-1}$.} 
    \label{tab:lambdacombo}
\end{table*}
\begin{table*}[]
    \centering
    \renewcommand{\arraystretch}{1.5}
    \tiny
    \setlength{\tabcolsep}{1pt}
    \begin{tabular}{c|c|c|c|c|c|c|c|c|c}
         & $a_0$ & $a_1$ & $a_2$ & $a_3$ & $a_4$ & $b_1$ & $c_1$ & $d_2$ & $e_2$ \\ \hline
         $M/R$ base fit & $1.8374\times10^{-1}$ & $-1.7316\times10^{0}$ & $7.6350\times10^{0}$ & $-1.7334\times10^{1}$ & $1.5597\times10^{1}$ & --------- & --------- & --------- & --------- \\ \hline
         $R_{1.4}$ terms only & $2.6392\times10^{-1}$ & $-2.2970\times10^{0}$ & $9.9112\times10^{0}$ & $-2.4162\times10^{1}$ & $2.3061\times10^{1}$ & $-4.7595\times10^{-3}$ & --------- & $1.7971\times10^{-2}$ & --------- \\ \hline
         $R_{1.4}$ and $\frac{R_{1.4}}{R_{1.8}}$ terms & $2.2483\times10^{-1}$ & $-2.2151\times10^{0}$ & $9.9363\times10^{0}$ & $-2.4251\times10^{1}$ & $2.3173\times10^{1}$ & $-4.1061\times10^{-3}$ & $3.0979\times10^{-2}$ & $1.6430\times10^{-2}$ & $-6.5181\times10^{-2}$ 
    \end{tabular}
    \caption{Best-fit coefficients for the $M/R$ fits to $R^4/(M^3\tau)$ with mixed second-order fit coefficients: the ``base" quartic fit in $M/R$, the quartic fit in $M/R$ plus a linear $R_{1.4}$ correction term and a $d_2R_{1.4}\frac{M}{R}$ term, and the quartic fit in $M/R$ plus linear correction terms in $R_{1.4}$ and $R_{1.4}/R_{1.8}$ and both $d_2R_{1.4}\frac{M}{R}$ and $e_2\frac{R_{1.4}}{R_{1.8}}\frac{M}{R}$ terms. The coefficients $a_n$, $c_1$, and $e_2$ are all dimensionless, whereas $b_1$ and $d_2$ have units of $\text{km}^{-1}$.} 
    \label{tab:compactextra}
\end{table*}

\end{document}